\documentclass[preprint,12pt]{elsarticle}

\usepackage{amsmath}
\usepackage{amssymb}
\usepackage{multirow}
\usepackage{xcolor}

\begin{document}

\begin{frontmatter}



\title{Data-Based Models for Hurricane Evolution Prediction: A Deep Learning Approach}



\author{Rikhi Bose, Adam Pintar \& Emil Simiu}

\address{Engineering Laboratory, National Institute of Standards \& Technology, 
            Gaithersburg, MD 20899, U.S.A.}

\begin{abstract}
Fast and accurate prediction of hurricane evolution from genesis onwards is needed to reduce loss of life and enhance community resilience. 
In this work, a novel model development methodology for predicting storm trajectory is proposed based on two classes of Recurrent Neural Networks (RNNs). 
The RNN models are trained on input features available in or derived from the HURDAT2 North Atlantic hurricane database maintained by the National Hurricane Center (NHC). 
The models use probabilities of storms passing through any location, computed from historical data. 
A detailed analysis of model forecasting error shows that Many-To-One prediction models are less accurate than Many-To-Many models owing to compounded error accumulation, with the exception of $6-hr$ predictions, for which the two types of model perform comparably. 
Application to 75 or more test storms in the North Atlantic basin showed that, for short-term forecasting up to 12 hours, the Many-to-Many RNN storm trajectory prediction models presented herein are significantly faster than ensemble models used by the NHC, while leading to errors of comparable magnitude. 
\end{abstract}



\begin{highlights}
\item We show that the use of $6-hr$ displacement probabilities as input features to Recurrent Neural Networks (RNNs) increases accuracy of the storm trajectory forecasting
\item We demonstrate compounded error accumulation in forecasts for the Many-To-One type RNN prediction models applied to storm trajectory forecasting. 
As a remedy, we highlight the use of the Many-To-Many type RNN prediction models. 
\item For short-term forecasting up to 12 hours, the Many-to-Many RNN storm trajectory prediction models presented herein are significantly faster than ensemble models used by the NHC, while leading to errors of comparable magnitude.
\end{highlights}


\end{frontmatter}


\section{Introduction}
\label{sec:intro}

Hurricanes and tropical storms are rotating storms originating in the Atlantic basin. 
The maximum sustained wind speeds exceed 74 $mph$ (33 $ms^{-1}$) for hurricanes, and are comprised between 39 $mph$ (17 $ms^{-1}$) and 74 $mph$ for tropical storms. 
The storms form as warm moist air rises from the sea surface and is replaced by cold dry air. 
This results in large cloud systems that rotate about their low-pressure core region owing to the Coriolis effect. 
In the northern hemisphere the rotation is counterclockwise. 
The probability of a storm formation is maximum during the months of highest sea surface temperature. 
The purpose of this paper is to present a simple Deep Learning (DL) based methodology for forecasting North Atlantic basin storm trajectories from genesis. 

The National Hurricane Center (NHC) uses several models for forecasting storm tracks and intensity. 
These models may be characterized as statistical, dynamical and statistical-dynamical. 
Statistical models (e.g., the Climatology and Persistence (CLIPER) model \citep{neumann1972alternate}) are typically less accurate than state-of-art dynamical models and are used as baseline models. 
Dynamical models, such as National Oceanic and Atmospheric Administration's (NOAA) Geophysical Fluid Dynamics Laboratory (GFDL) hurricane prediction model \citep{kurihara1995improvements} numerically solve thermodynamics and fluid dynamics equations that govern the atmospheric motions and may require hours of modern supercomputer time for providing even a $6–hr$ forecast. 
Other models combine the skills of the statistical and dynamical models. 
To take advantage of the skills of available statistical and dynamical models NHC uses ensemble prediction systems (EPSs) (such as the ensemble model of the European Centre for Medium-Range Weather Forecasts (ECMWF) \cite{buizza1999stochastic, richardson2000skill}) that can be more accurate and reliable than their components. 

DL methods have recently taken giant strides in using large amounts of data to make accurate predictions in fields such as computer vision that have traditionally been very challenging for conventional statistical models. 
We propose a DL approach for efficient prediction of hurricane trajectories over several $6-hr$ time intervals. 
The target variables in the DL hurricane track prediction problem are the coordinates of a storm's center. 
The DL models include as inputs storm features available in or derived from the hurricane database maintained by the NHC. 

Such a data-based approaches have been recently applied to several aspects of weather forecasting \citep{schultz2021can}. 
Hurricane forecasting using data-based approaches have mainly been formulated as image processing problems by \citet{giffard2018deep, kim2019deep, ruttgers2019prediction, chen2019hybrid, chen2020novel}. 
\citet{giffard2018deep} used atmospheric flow velocity and pressure information at several heights above the Earth's surface to train a model comprised of several convolutional and feed-forward neural network layers; a mean error of 115 $km$ for $24-hr$ forecasts was obtained for test data over the North Atlantic.  
A Convolutional Long short-term memory (ConvLSTM) model, was used by \citet{kim2019deep} to extract spatio-temporal information from a large database of instantaneous atmospheric conditions recorded as a pixel-level history of storm tracks. 
The ConvLSTM comprises Convolutional Neural Network (CNN) layers and layers of a class of RNN, called Long short-term memory (LSTM) \cite{hochreiter1997long}. 
A tensor-based Convolutional Neural Network (TCNN) was used by \citet{chen2020novel} to improve hurricane intensity forecasts in coupled TCNN (C-TCNN) and Tucker TCNN (T-TCNN). 
\citet{ruttgers2019prediction} utilized the Generative Adversarial Network (GAN) to forecast storm trajectories using satellite images. 
They used image time-series of typhoons in the Korean peninsula for model training. 
The GAN model was tested on 10 storms previously unseen by the model during training. 
The average prediction error for $6-hr$ forecasts for the test storms was 95.6 km. 
More recently, motivated by the work of \citet{giffard2018deep}, \citet{boussioux2020hurricane} developed storm forecasting models utilizing reanalysis data extracted during tropical storms. 
Several DL models, such as CNNs, GRUs, Transformers \citep{vaswani2017attention}, as well as ML models, e.g., XGBoost \citep{chen2016xgboost}, were utilized. 
They have reported a minimum mean $24-hr$ forecast error of $\sim110$ $km$ for test storms between 2016 and 2019 in the Atlantic basin. 
However, the aforementioned models are data-hungry and in most instances require current satellite images, or in some instances atmospheric post-storm reanalysis data that are not instantaneously available, and may be affected by non-negligible errors themselves; for those reasons the use of such data is a disadvantage in a forecasting context. 
This was noted by some of the aforementioned researchers themselves. 
\citet{boussioux2020hurricane} in their paper comment, `\textit{… although our models can compute forecasts in seconds, the dependence on reanalysis maps can be a bottleneck in real-time forecasting}'.

In \cite{moradi2016sparse}, a flexible sparse RNN architecture was used to model the evolution of North Atlantic hurricane tracks. 
Available time records of the target storm were initially compared with those of other storms, and a Dynamic Time Warping technique was used to assign similarity scores to each storm available in a historical storm database. 
Only storms with high similarity scores were chosen for model training. 
Storm evolution prediction was thus improved by incorporating historical trends. 
However, for obtaining appropriate storms for the model to train on, significant numbers of time records from the target storm are required. 
In the present work, we account for translation trends of historical storms by using storm displacement probabilities as input features. 
The storm displacement probabilities take into consideration $6-hr$ displacement of all storms from a given computational cell to adjacent cells in a coordinate-transformed domain. 

In another attempt to use RNN models, authors in \cite{alemany2019predicting} used a grid-based Neural Network approach for hurricane trajectory forecasting. 
A Long short-term memory (LSTM) RNN model predicted storm-wise scaled grid numbers in the $2-D$ latitude-longitude domain. 
The grid-based prediction scheme was based on the fact that, for a historical storm, the distance being traveled is proportional to the number of time records available for it. 
The model forecasts a storm's location six hours in advance, and is claimed to improve upon the forecasting performance of the models developed in \cite{moradi2016sparse}. 
They used storm-wise data scaling, which renders predictions for new storms impossible. 
Moreover, no results were provided for long-term forecasts. 
The compounded error accumulation, a most pressing issue related to the application of NNs for storm track prediction, was not considered in their work. 
In the present effort, we use LSTM-RNNs to provide long-term forecasts for storm evolution up to thirty hours in advance. 
We demonstrate the compounded error accumulation problem and a remedy for it using Many-To-Many prediction-type RNNs. 

In a more recent work, CNNs coupled with Gated recurrent unit (GRU) RNNs were employed by \citet{lian2020novel}. 
The use of a feature selection layer before the NN layers augmented the model's learning capability from the underlying spatial and temporal structures inherent in trajectories of tropical cyclones. 
Their model utilized both CNNs and GRUs. 
They did not segregate the training, validation and test storms; 6108 data sequences were used for testing the model. 
The model's performance was compared with the performance of other numerical and statistical weather forecasting models. 
The NN model used in \cite{lian2020novel} was more accurate than the statistical model used in \cite{jeffries1993tropical} and a numerical model used in \cite{demaria1992nested} for long-term forecasts. 
In particular, the $72-hr$ track forecast error was was less than half the errors reported for traditional track forecasting models. 
However, they reported a $12-hr$ forecast error of $\approx 100$ km (62 mi) between the predicted and true cyclone-eye locations, which is comparable or marginally less than the errors incurred by the aforementioned statistical and dynamical models. 
The RNN models developed herein were tested for hundreds of validation and test storms, and the prediction errors were extensively analyzed. 
Computed over different splits of the HURDAT2 database, mean $6-$ and $12-hr$ forecast errors for the prediction models were $\sim 33$ $km$ and 72 $km$, respectively. 
To our knowledge, the models outperform all data-based models for short-term trajectory forecasting up to 12 hours. 

Despite the substantial existing work on the topic, there exist opportunities for improvement, specifically in the context of formulating a model usable in real-time that uses readily available information. 
In this regard, the use of historical storm trajectory information is the most suitable as these are easily avilable. 
We find herein that the use of $6-hr$ displacement probabilities to have predictive power. 
The storm displacement probabilities embed historical storm translation trends by considering the $6-hr$ displacements of all storms from a given location. 
Results from the RNN-DL models presented herein demonstrate that the use of this simple input variable is able to provide results that are competitive in accuracy with the state-of-the-art models using large volumes of atmospheric flow field/ satellite image data. 
The flexible model forecasting scheme can predict trajectories even when only one record is available, i.e., from genesis, unlike most ML/ DL based models developed so far. 
Furthermore, the current work emphasizes that traditional ML/ DL accuracy measures are not applicable to storm trajectory forecasting, especially for long-term forecasts that are prone to compounded error accumulation. 
The error analysis of the presented models highlight this pressing issue and suggests the use of Many-To-Many type prediction architectures in RNNs that can defer the compounded error accumulation.

The paper is structured as follows. 
In Section \ref{sec2}, the database used for model development, is discussed from the statistical, the feature engineering, and the model formulation points of view. 
The methodology used for the calculation of $6-hr$ storm displacement probabilities is also described there. 
Section~\ref{sec3} discusses the model type, its architecture and implementation, training strategies, and hyperparameter tuning. 
Section~\ref{sec4} contains the prediction results of the trained models and an extensive analysis of forecasting error. 
Section~\ref{sec4} also compares the predicted trajectories with trajectories of historical test storms, and discusses limitations of the models developed herein and the scope for future improvements. 
Conclusions are drawn in Section \ref{sec5}.  

\section{Database \& Feature engineering}
\label{sec2}

\begin{figure*}[!ht]
    \begin{center}
        \includegraphics[trim=3.0cm 1cm 2.0cm 1cm,clip,width= 0.9\linewidth]{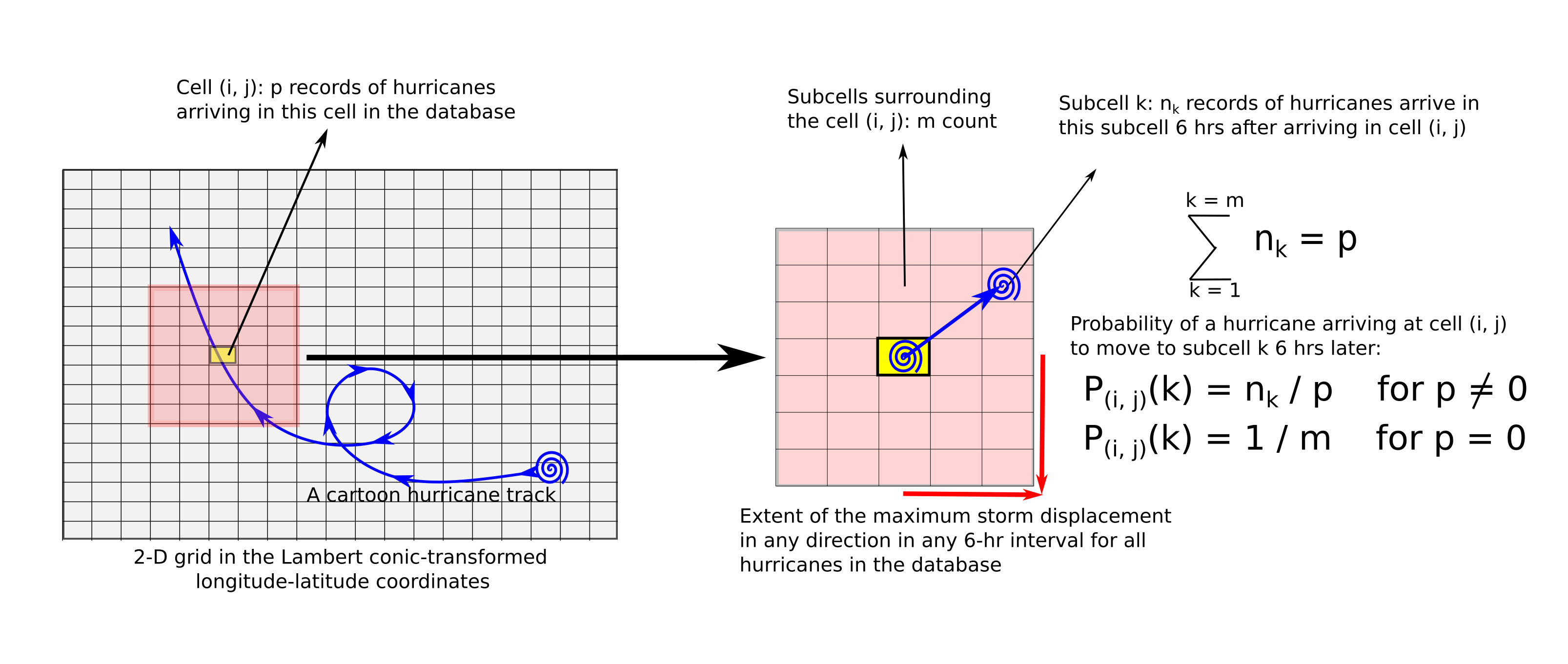}
        \caption{Schematic depicting the calculation of displacement probabilities associated with each cell in the grid domain.}
        \label{f2}
    \end{center}
\end{figure*}

\subsection{Feature space} 

The NHC conducts post-storm analyses and maintains the HURDAT2 database \cite{jarvinen1984tropical, landsea2013atlantic} that lists features of Atlantic-basin storms at $6-hr$ intervals (e.g., date and time, landfall/ intensity status, latitude and longitude, maximum $1-min$ wind speed at 10 $m$ elevation, central pressure). 
However, only part of the database is useable for DL purposes. 
The present models use a maximum of five time records as input. 
Of the 1893 listed storms, 1736 storms have at least seven time records with maximum wind speeds listed for all records. 
Only 560 of these have both the central pressure and maximum wind speed included for all records. 
Based on physical reasons, the central pressure is highly correlated with the maximum wind speed; in fact, observed maximum wind speed varies approximately linearly with the central pressure with a negative slope for all time records of the aforementioned 560 storms in HURDAT2. 
Therefore, the central pressure was excluded from the input feature space which allowed us to use a much larger volume of data without losing much useful predictive information. 
We used the latitude ($\phi$), longitude ($\lambda$) and maximum wind speed ($w_m$) to construct the input feature space. 

To reduce the effect of areal distortion associated with the spherical coordinate system at higher latitudes, a transformed pair of $x-$ and $y-$coordinates obtained by Lambert's conic conformal (LCC) projection \cite{lambert1972notes} was used instead of $\phi$ and $\lambda$. 
LCC projection has superior properties in mid-latitudes. 
Two standard parallels associated with the conic projection ($\phi$'s where the cone passes through the sphere) used in the current transformation are $33^\circ N$ and $45^\circ N$. 
Most of the U.S. mainland is contained within these standard parallels. 
No distortion occurs along the standard parallels but distortion increases away from these. 
In addition, we use the storm translation direction and speed as inputs, as these are important for storm intensity modeling \citep{schwerdt1979meteorological, emanuel2006statistical}. 
Six-hour-averaged translation speed ($V$) and direction ($\theta$) are calculated going backward in time. 
The Haversine formula gives the distance ($d$) between storm locations at two consecutive time instants.  

\begin{eqnarray}
V_l = \frac{d(\phi_{l-1}, \phi_l, \lambda_{l-1}, \lambda_l)}{\Delta t \equiv 6 hrs} \\
\theta_l = \tan^{-1}\bigg( \frac{\phi_l -\phi_{l-1}}{\lambda_l -\lambda_{l-1}} \bigg)
\end{eqnarray}

In HURDAT2, $w_m$ is approximated to the nearest 10 $kt$ (5.14 $ms^{-1}$) between 1851 and 1885 and to the nearest 5 $kt$ (2.57 $ms^{-1}$) thereafter. 
Binned $w_m$ is used from the HURDAT2 database. 
Both $w_m$ and $V$ were expressed in $ms^{-1}$ units. 

It is necessary for the input features to contain information associated with trends of past storm motion \cite{moradi2016sparse, dispProb}. 
Those trends are encapsulated in $6-hr$ storm displacement probabilities (d.p.) computed from the historical storms, that are included as input features at each time instant. 
A schematic depicting the calculation of d.p. is shown in Figure \ref{f2}. 
At first, the $2-D$ domain bounded by the extents of the $x-$ and $y-$ coordinates (the extents are the maximum and minimum of each coordinate corresponding to any storm record in the considered database) was decomposed into rectangular computational cells. 
Consider all storms for which a record is contained in any one of those computational cells. 
For those storms, the maximum number of cells traversed by any storm in any $6-hr$ interval in the $x-$ and $y-$ directions are $m_x$ and $m_y$, respectively. 
Therefore, associated with any given cell $(i, j)$ (colored yellow in Figure \ref{f2}) is a set of $m = (2m_x + 1)(2m_y + 1)$ cells within which all $6-hr$ displacements of any historical storm is contained. 
Such a set is colored red in Figure \ref{f2}. 
Assume, that there are $p$ records of hurricanes that arrived at the $(i, j)-th$ cell and then transitioned to the $k^{th}$ associated cell in the next 6 hours. 
If this leads to $n_k$ records being sampled at $k^{th}$ associated cell, $\sum_1^m n_k = p$. 
Therefore, the displacement probability of a storm arriving at the $(i, j)-th$ cell and transitioning to the $k^{th}$ associated cell in the next 6 hours is, 

\begin{equation}
    p_{(i,j)}(k) =   
    \begin{cases}
        \frac{n_k}{p},& \text{if } p\neq 0\\
        \frac{1}{m},& \text{if } p = 0
    \end{cases}
\end{equation} 

For a fine grid the number of cells with $p\simeq0$ would be larger than for a coarse grid. 
For a finer grid (and/ or for a very thin upper tail of the distribution of $V$), $m$ would increase, and $n_k$ would decrease. 
In addition to an increased number of features at each instant, d.p. from a cell to adjacent cells could be more biased on a specific historical storm's displacement, which is not desirable for the prediction of new storms. 
On the other hand, if the grid is too coarse, storm motion trends reflected by the d.p. could be obscured. 


\begin{figure*}[!ht]
    \begin{center}
        \includegraphics[trim=0 0 0 0,clip,width=0.45\textwidth]{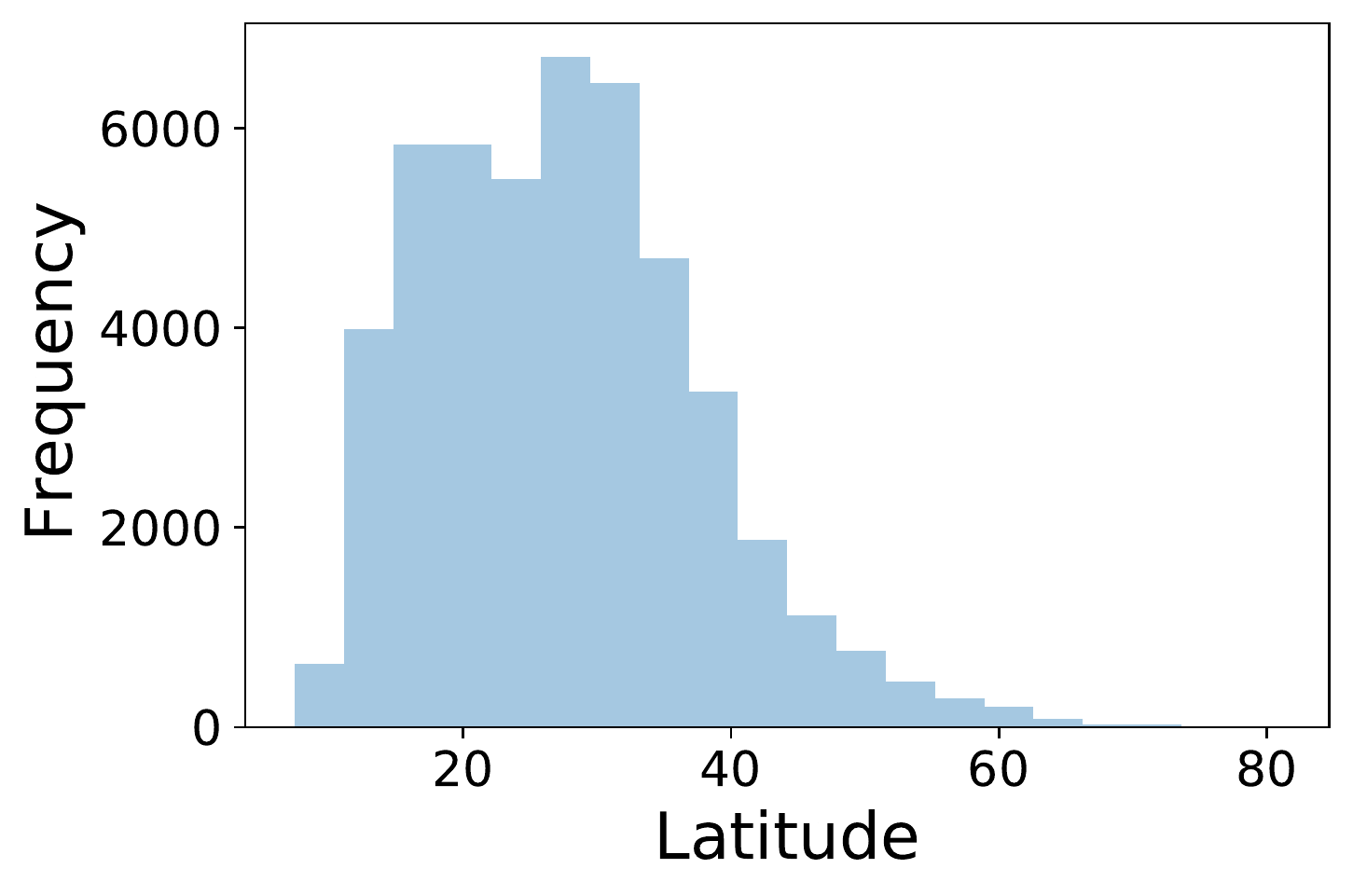}
        \includegraphics[trim=0 0 0 0,clip,width=0.45\textwidth]{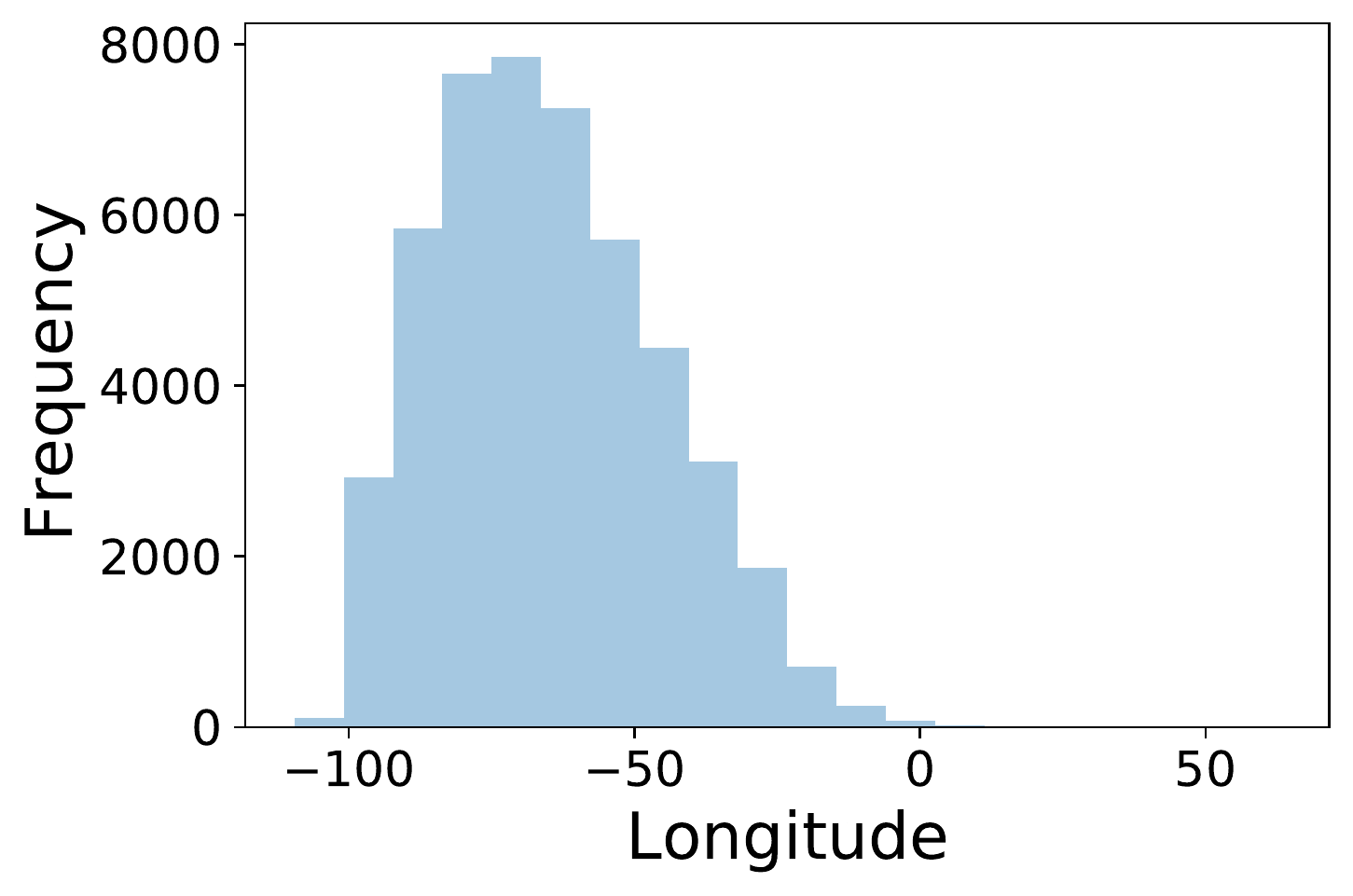}
        \includegraphics[trim=0 0 0 0,clip,width=0.45\textwidth]{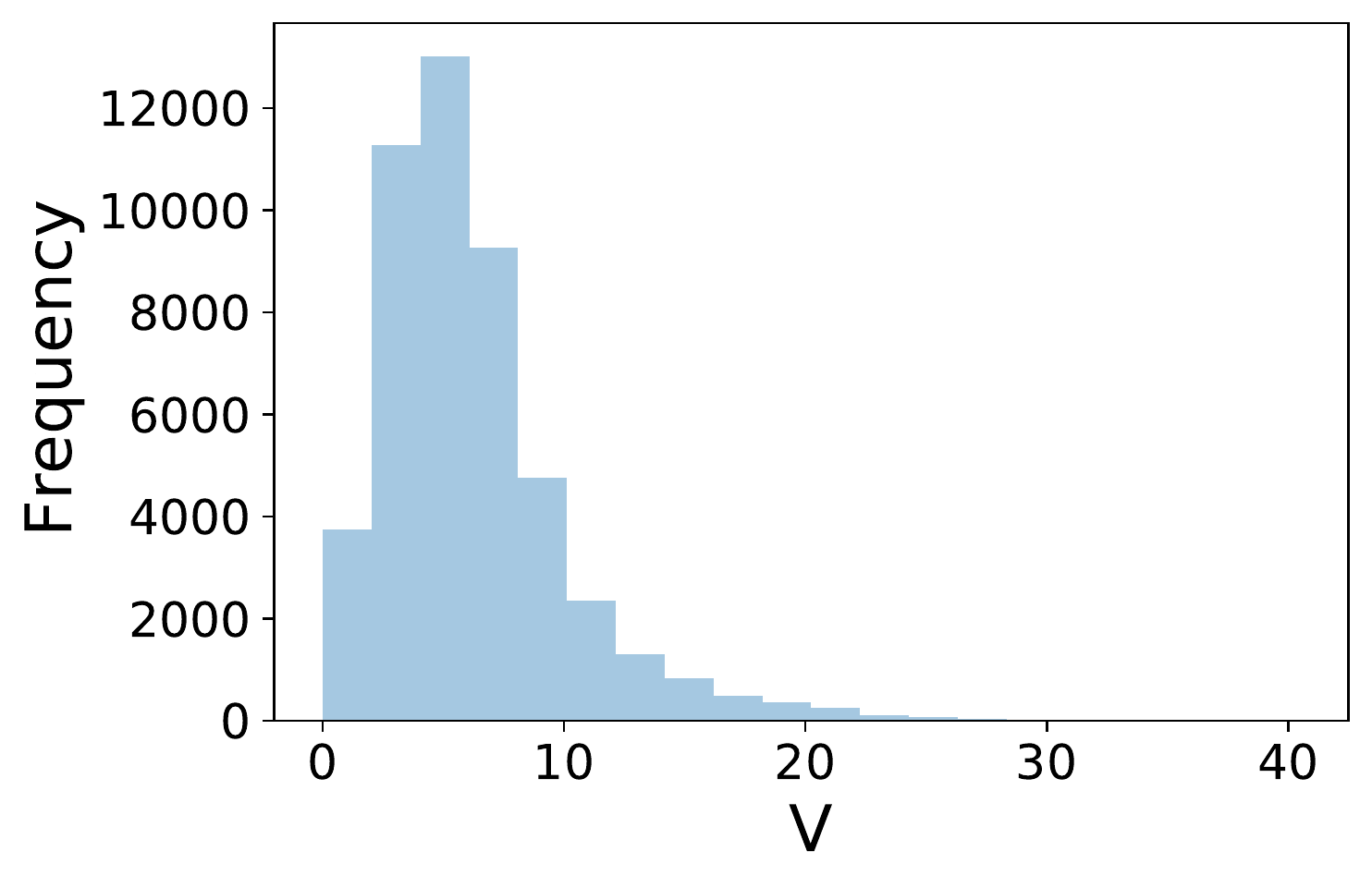}
        \caption{Histograms of latitude ($\phi$) and longitude ($\lambda$) in degrees and storm translation speed ($V$) in $ms^{-1}$. 
        Positive values of $\phi$ and $\lambda$ represent the northern and eastern hemispheres, repectively. 
        1736 storms from the $HURDAT2$ database are used (see text). 
        }
        \label{fPDF}
    \end{center}
\end{figure*} 

\subsection{Statistics} The computational domain containing the 1736 storms with at least seven records, including $w_m$, covers $\phi \in [7.4^\circ N, 81^\circ N]$ and $\lambda \in [109.5^\circ W, 63^\circ E]$. 
The storm translation speed, which does not exceed $V = 40.5$ $ms^{-1}$ for any storm in the database, determines the number $m$ of subcells associated with a given computational cell as shown in Figure \ref{f2}. 
Histograms of $\phi$, $\lambda$ and $V$ are shown in Figure \ref{fPDF}. 
Only for 33 of the 47880 time records was $\phi > 70^\circ N$, and only for 26 records was $\lambda > 10^\circ E$. 
Only 93 records of storm motion satisfied the condition, $V > 25$ $ms^{-1}$. 
Therefore, the 71 storms violating these limits were excluded; 1665 storms remained in the final database used for model development. 
The coordinate limit $\phi = 70^\circ N$ is chosen because it is even north of the Arctic circle. 
The limit for longitude, $\lambda = 10^\circ E$ ensures that storm records from all of the Atlantic ocean in the Northern hemisphere upto the western shores of the continent of Africa is retained in the database. 
We chose $V \le 25$ $ms^{-1}$ to limit the magnitude of $m$.

Storm d.p. were calculated based on the evolution of these 1665 storms. 
61 grid points were used in both $x$ and $y$ directions, resulting in 3600 computational cells decomposing the whole 2-D LCC projected domain. 
Despite application of the three above mentioned criteria, for 1687 out of the 3600 computational cells, $p=0$. 
The maximum $6-hr$ displacement for any storm in the 1665 storm-database could be captured with $m_x=m_y=4$. 
So, the d.p.s were calculated for $m=9\times9=81$ surrounding cells for each computational cell. 
Consequently, the number of input features for each time record was 86 (LCC transformed $x$ and $y$ coordinates, $V$ and $\theta$, $w_m$, and historical $6-hr$ d.p. calculated at 81 subcells associated to the cell containing $x$ and $y$).

\section{Model development}
\label{sec3}

Once a database was obtained, we proceeded to the model development. 
Both classification and regression problems may be formulated. 
In the current work, only the regression problem is considered for which we used the LSTM RNN \cite{hochreiter1997long}. 

In the supervised regression problem being considered, inputs to the models are the features at the chosen number of input time instants, $n$ ($86 \times n$  for each prediction). 
The model outputs are the storm position(s) $x$ and $y$ at $6-hr$ intervals for the chosen number of output timesteps. 
The model performance is measured by the loss function defined as the mean squared error ($m.s.e.$) between predicted and true position(s) of $N$ samples (here, $N$ is the number of data sequences used from the storms in the database), $$m.s.e. = \frac{\sum_{i=1}^N(x_{predicted, i} - x_{true, i})^2+(y_{predicted, i} - y_{true, i})^2}{N}.$$ 
Based on the gradients of the $m.s.e.$, model weights are updated so that the $m.s.e.$ is minimized. 

\subsection{Model architecture \& Implementation} 

RNNs extract pattern and context from sequences. 
As each storm in the HURDAT2 is a sequence of time records, RNNs are suitable. 
The LSTM architecture was prescribed \citep{hochreiter1997long} to tackle the vanishing gradient problem in RNN models for long sequences. 
In an LSTM unit, past information may be retained via a cell state that passes through all LSTM layers \citep{karpathy2015visualizing}. 

Models trained on varying numbers of input time records ($n$) can predict one or several time steps at once based on their architecture. 
Both Many-To-One ($M2O$) and Many-To-Many ($M2M$) type prediction algorithms have been used. 
Upon processing a sequence with $n$ time records, the $M2O$ models forecast the storm locations at one time instant, while the $M2M$ models used here output $n$ time records. 
From here onwards, models are named as $M2On$ or $M2Mn$ to reflect their architecture. 
As each timestep in the training database $\equiv6-hr$, the $M2On$ and $M2Mn$ models forecast 6 and $6n$ hours at once, respectively. 

In addition to the input and output layers, the $M2On$ models presented here have 3 layers of LSTM units. 
The numbers of neurons in the three LSTM layers were 128, 32 and 8, respectively. 
In preliminary tests, the performance of converged models did not improve with additional LSTM layers. 
Two bi-directional LSTM layers were used at either side of a repeat-vector layer for the $M2Mn$ models. 
Each LSTM cell in a bi-directional layer had an input and output dimension of 64; so 128 neurons were used in each layer. 
The numbers of trainable parameters for all the tested models were between 3 to 6 times the number of data sequences used for training. 

To increase model robustness, dropout is used to reduce a model's bias towards the training data. 
A dropout value of 0.1 was used for each hidden layer. 
The Adam algorithm \citep{kingma2014adam} was used for model optimization. 
Finally, the models were implemented in the Keras API with TensorFlow as the backend library. 
Further detail about the model architectures used herein, their implementation and additional considerations for their training is provided in \cite{bose2021forecasting}.


\subsection{Number of input time records ($n$)} An important consideration is the number of input time instants, $n$ that could minimize the model prediction error. 
Intuitively, increasing $n$ should make the model predictions more accurate. 
However, high values of $n$ imply less available time for preparation of a possible landfall. 
Also, the value of $n$ implicitly determines the number of data sequences available from the storm database for model training/ validation/ testing. 
Preprocessing, such as zero padding may be used at early stages of a storm's life span (available time instants $< n$). 
However, training of a model with zero-padded data is inefficient, because the mode of the number of records for all storms in the HURDAT2 database is 13, while storms with a very long lifespan, e.g., as long as $\sim 80$ time records, are relatively rare. 
If one chose to use zero padding so that each storm have, for example, 50 records, a significant number of time records fed to the network for training would be zeros. 
On the other hand, requiring 12 time records in each input sequence, for example, renders the zero padding unnecessary as only a few storms have a lifespan shorter than 13 time records in the database (12 input records + 1 output). 
Both $M2On$ and $M2Mn$ LSTM models were developed for a range of $n$, between 1 and up to 5 time instants ($n_{max} = 5$). 

\subsection{Data scaling} NNs perform better when the input data is scaled to the interval $\in [0, 1]$. 
Normalization of the whole database was preferred over a storm-wise feature scaling approach \citep{alemany2019predicting}. 
The storm-wise data scaling renders prediction of a new storm impossible, because the relevant scaling parameters for a feature for an entirely new storm are unknown a priori. 
Each feature $f$ is scaled using the Min-Max scaler as, $f_s = \frac{f - f_{min}}{f_{max}-f_{min}}$, so that the scaled feature $f_s \in [0, 1]$. 
This scaling is preferred over standardization, because, the d.p. also belong in this interval. 
In contrast to a storm-wise scaling approach, feature scaling for an entirely new storm is well defined. 

\subsection{Sequence generation} 

To generate a sequence for the $M2M5$ model, the life span of a storm must be at least 10 time-record long. 
1514 out of the 1665 storm database with at least 10 time records were used to facilitate comparison between the $M2On$ and $M2Mn$ models, which must be trained, validated and tested on the same set of storms. 

To avoid training and testing on the same storm, the 1665 storms were partitioned into training, validation, and testing sets. 
Test storms (5\% of the total) always contained a set of five important historical storms chosen a priori because of their destructiveness upon landfall and of the complexity of their trajectories. 
The rest of the storms were chosen randomly from the database without replacement. 
Of the 1665 storms considered, $\sim 80\%$ were used for training the model while 15\% storms were used for validation. 
Data sequences for training, validation and testing were generated separately from the segregated lists of storms. 
The training data always contained $\sim 24,000$ sequences or more. 

\subsection{Learning rate \& Batch size} 

In Keras, the default learning rate for the Adam optimizer is 0.001. 
At this value, the validation loss was prone to sudden jumps in and around the optimal valley's minimum. 
We reduced the learning rate by an order of magnitude every 250 epochs from an initial learning rate of 0.0001 until the models stopped improving. 

We used a batch size of 32 which significantly lowered training time in the Graphic Processing Unit (GPU) clusters. 
A model checkpoint was used after completion of each epoch to compute the validation error and store the weights in case the error obtained is the minimum at that point. 
At the end of model training, model weights yield the lowest loss function value for the validation sequences. 

\section{Results}
\label{sec4}

In the following discussion, models have been named $M2On$ and $M2Mn$, where, $n = 1, 2, 3, 4$ and 5 represents the number of records a model takes in as input. 
The $M2On$ and $M2Mn$ models forecast 6 and $6n$ hours at once, respectively. 

\begin{figure}[!ht]
    \begin{center}
        \includegraphics[trim=0cm 0cm 0cm 0cm,clip,width=0.4\textwidth]{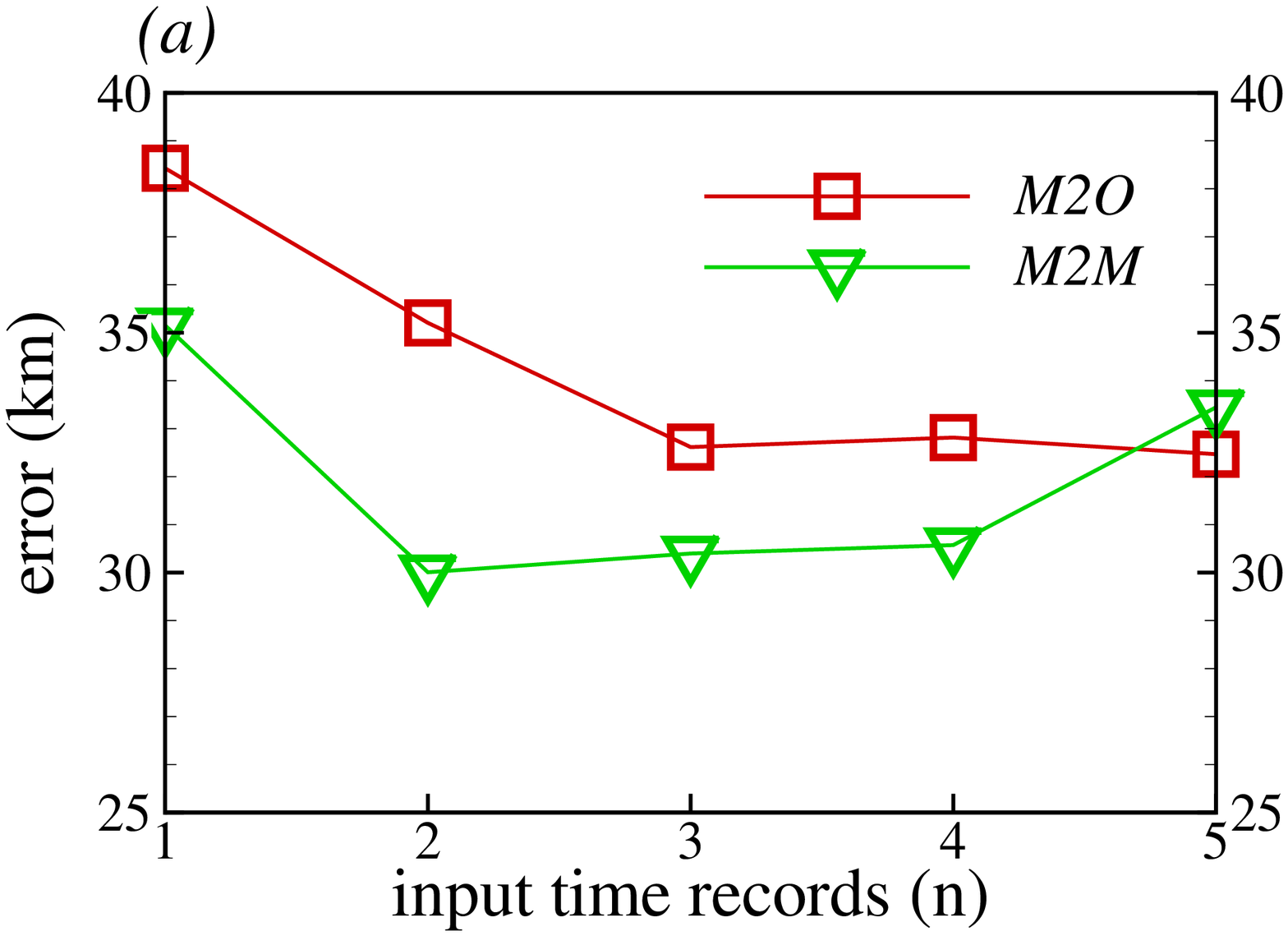}
        \includegraphics[trim=0cm 0cm 0cm 0cm,clip,width=0.4\textwidth]{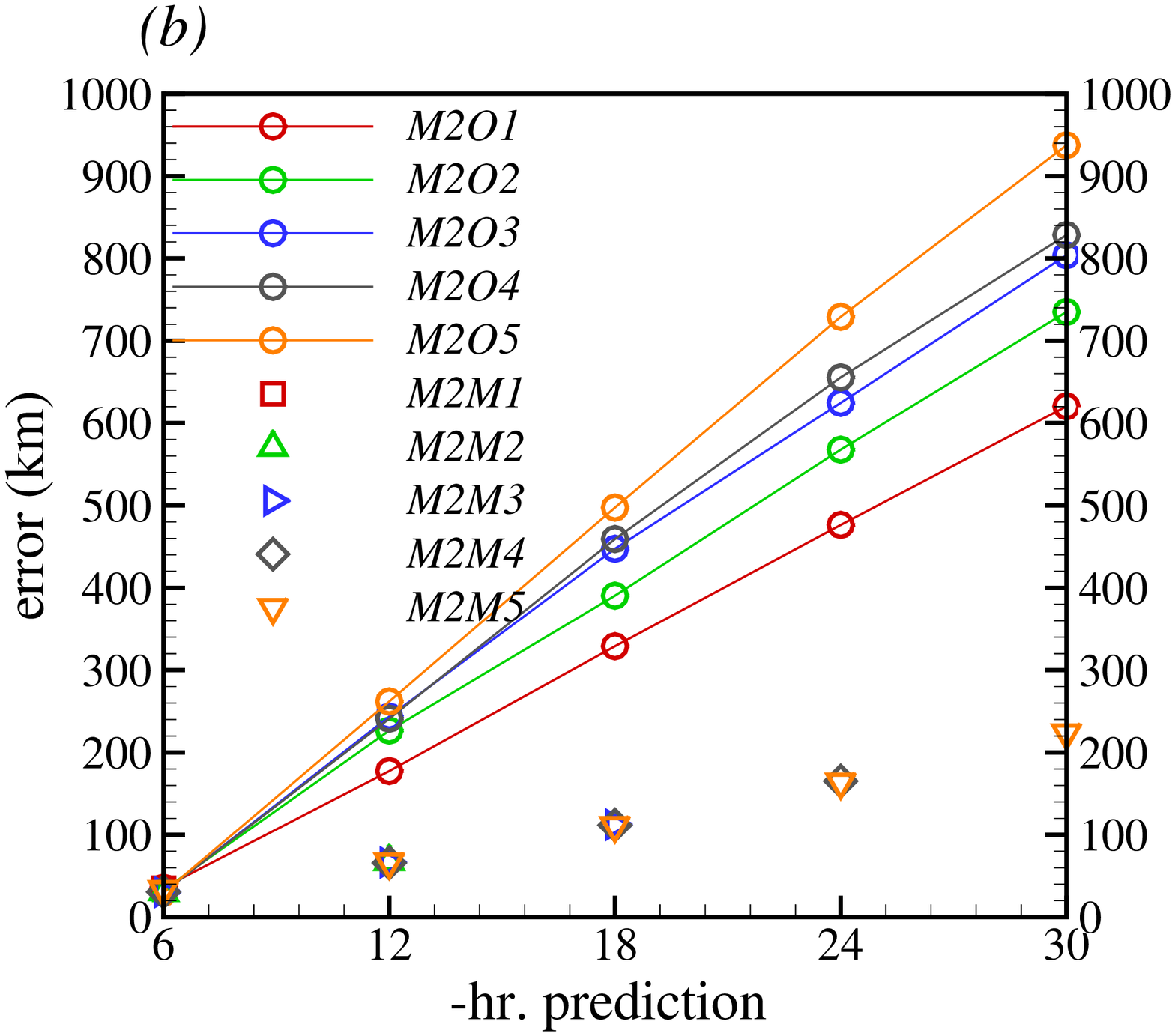}
        \caption{
        Mean ($a$) $6-hr$ forecast error and ($b$) 6, 12, 18, 24 and 30 --$hr$ forecast errors in distance computed on test storms for $M2O$ and $M2M$ models trained on different number of input time records. 
        In ($b$), errors from the $M2O$ models are represented by circular symbols and the $M2M$ models are represented by other types of symbols; models trained on a given number of time records are represented by symbols of same color. 
        All models were trained, validated and tested on the same set of storms.  
        }
        \label{f6hrErr}
    \end{center}
\end{figure} 

\subsection{Forecast error: Compounded error accumulation} 


\begin{figure}[!ht]
    \begin{center}
        \includegraphics[trim=0cm 0cm 0.0cm 0cm,clip,width=0.495\textwidth]{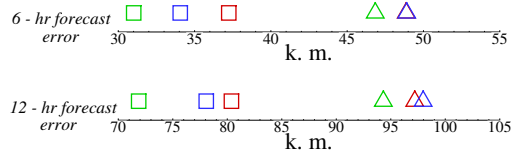}
        \caption{Mean $6-hr$ (top) and $12-hr$ forecast errors for the $M2M2$ models trained on input features including $6-hr$ displacement probabilities ($\square$) and excluding those ($\Delta$). 
        The test storms are in different time intervals: 2005--2009 (red), 2011--2015 (green), and 2016--2019 (blue). 
        }
        \label{fDP}
    \end{center}
\end{figure}

The models were tested on 228 validation and 75 test storms to compute the average error in distance between the predicted and the true positions of a storm’s eye at each prediction step. 
To facilitate comparison, for the results presented herein, both $M2On$ and $M2Mn$ models were trained, validated and tested on the same set of storms. 
Average $6-hr$ and $6n-hr$ forecast errors for the test sequences are shown in Figure \ref{f6hrErr} for all models. 
The errors for the validation sequences are very similar and are not reported herein. 
Figure~\ref{f6hrErr}($a$) shows the mean $6-hr$ forecast error plotted for different input time records ($n$). 
Overall, the $M2M$ models are more accurate. 
The $M2M2$ model performs best of all models for $6-hr$ forecasting; mean error computed for the validation and test storms are 34.2 and 30 km, respectively. 
This is in accordance with \citet{emanuel2006statistical} who noted a near linear statistical relationship between time-rate changes in translation variables at current and previous timesteps implying that the use of two time records may be optimal $6-hr$ forecasting. 
Although the number of trainable parameters for all models is similar, the $M2On$ models only slightly improve with increase in $n$. It must be noted that the mean errors shown are for only a single split of the storm database. 
The $M2Mn$ model predictions worsen slightly with increase in $n > 2$. 
However, a maximum difference of 5 $km$ in mean $6-hr$ forecast error between the $M2On$ and $M2Mn$ models is obtained for $n=2$. 

For a given number of available time records for a storm, which model predicts a storm's future trajectory more accurately over longer time intervals? 
Figure~\ref{f6hrErr}($b$) reports the mean forecasting errors for up to 30 hours computed for the same test storms as in Figure \ref{f6hrErr}($a$). 
Although the $M2Mn$ models can forecast $n$ time records (i.e., $6n$ hours) at once, the $M2On$ models however predict only one time record. 
To obtain long-term forecasts with the $M2On$ models, the predicted record must be used as an input for the next time instant and so on. 
The forecasting errors for up to 5 time steps in advance obtained in this manner are also reported for all $M2O$ models. 
Although the LCC projected $x$-- and $y$-- coordinates, $V$ and $\theta$ and the $6-hr$ storm d.p. are updated at each time step, the true value of $w_m$ is used from the database. 
This provides a fair comparison between $M2O$ and $M2M$ model accuracies in predicting several future time steps. 
This prediction scenario is similar to a real-time storm trajectory forecast.

Due to the aforementioned time record updating procedure for the $M2O$ models, the 12, 18, 24 and 30--$hr$ forecasts are prone to compounded error growth. 
Error incurred at each prediction step worsens the forecasting accuracy at future steps as the predicted coordinates which deviate from the truth are used as input for predicting the future storm locations. 
Due to the error accumulation, the $M2O$ model forecasts beyond the $1^{st}$ prediction step worsen for increasing $n$, as for a higher $n$ more erroneous input time records are fed to the model for forecasting. 
All $M2M$ models have similar accuracy over each $6-hr$ interval. 
Therefore, given an initial time sequence to predict a storm's evolution, especially to predict it several hours in advance, $M2M$ models are more reliable. 
The average $12-hr$ ($24-hr$) forecasting error for test sequences for the $M2M4$ and $M2M5$ models are 65.9 and 66.85 $km$ (165.2 and 163.9 $km$), respectively. 

\subsection{Displacement probabilities}

The advantage of using the $6-hr$ d.p. as input features is demonstrated by reporting mean errors in distance for the $M2M2$ model trained on input features including, and excluding, the $6-hr$ d.p. 
Three different splits of the database for train, validation and test storms are used which have been also used later herein for benchmarking. 
The NHC produces forecasting error estimates for specified time periods, e.g., for the periods 2005-2009, 2011-2015, and 2016-2019. 
Each of the three test databases used includes all storms with at least 10 time records from either of these periods. 
The remainder of the 75 storms are chosen randomly without replacement. 
However, the same sets of training, validation and test storms are used for training models with/ out d.p. as inputs. 
The results are shown in Figure \ref{fDP}. 
Use of the d.p. is clearly advantageous and significantly improves model performance. 
The maximum gain in accuracy is obtained for the period 2011-2015; using d.p., mean $6-hr$ and $12-hr$ forecasting errors are reduced by $\sim16$ $km$ and 23 $km$, respectively. 
The accuracy gain for the chosen test datasets is at least $\sim11$ $km$ for $6-hr$ forecasts and 17 $km$ for $12-hr$ forecasts.

\begin{figure*}[!ht]
    \begin{center}
        \includegraphics[trim=2cm 0cm 2cm 0cm,clip,width=0.45\textwidth]{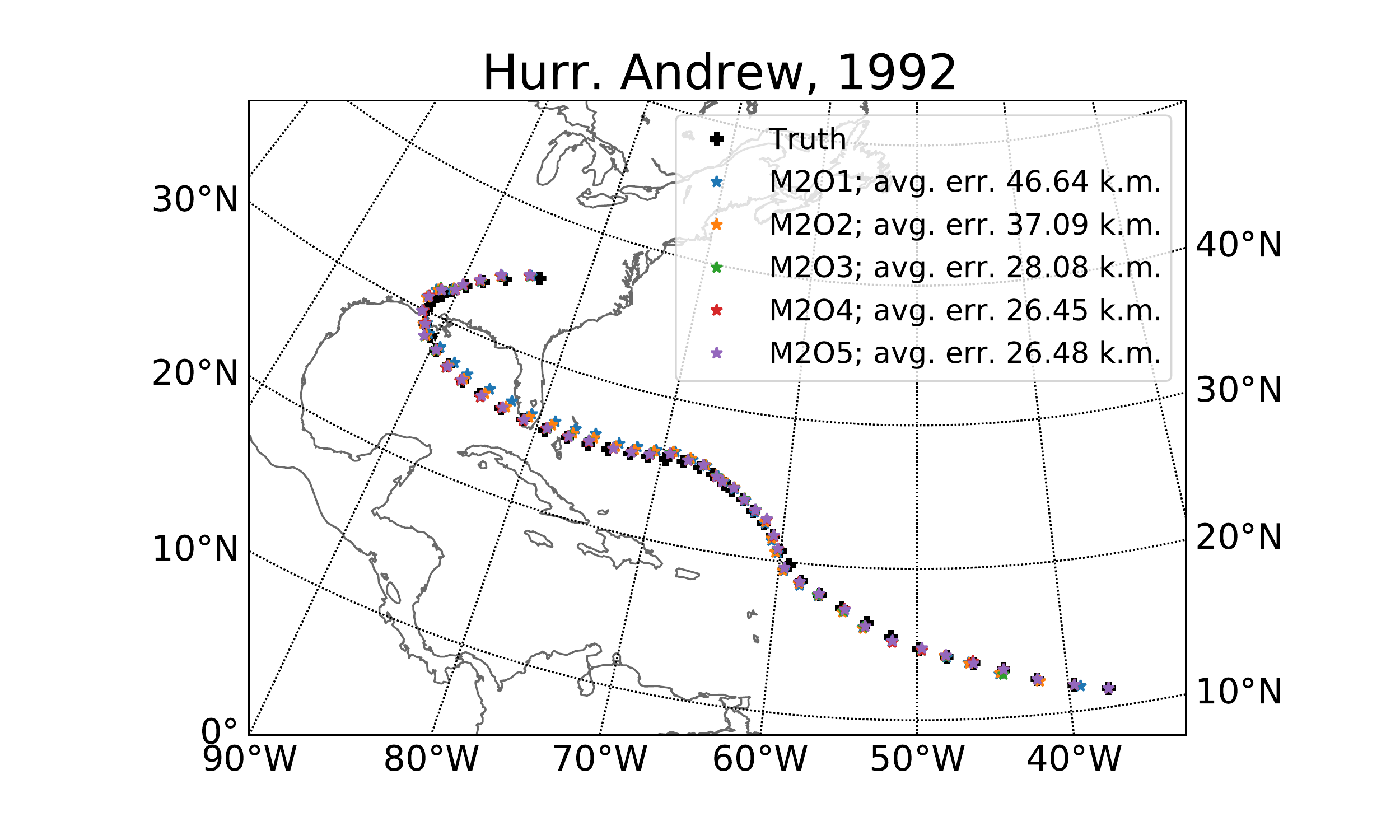}
        \includegraphics[trim=2cm 0cm 2cm 0cm,clip,width=0.45\textwidth]{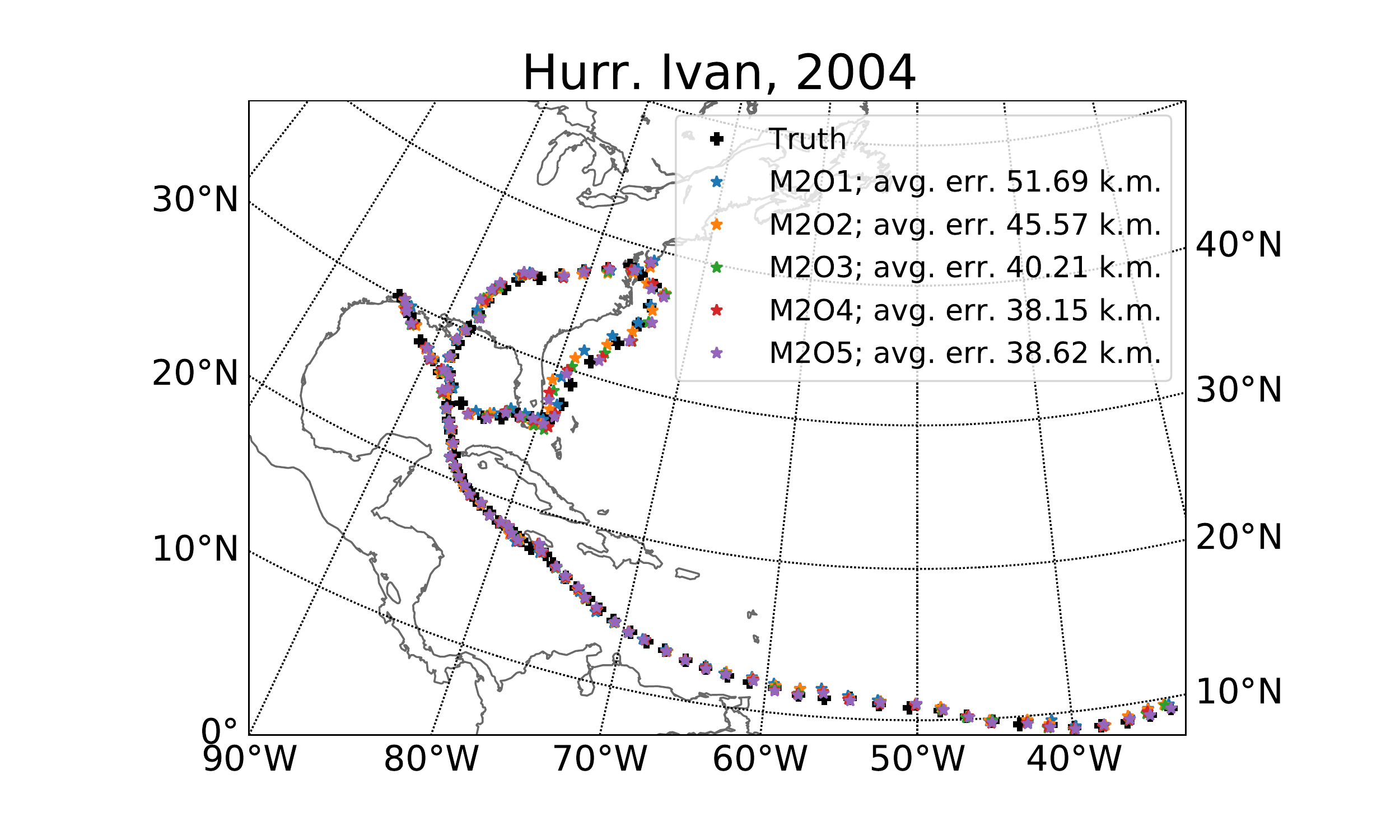}
        \includegraphics[trim=2cm 0cm 2cm 0cm,clip,width=0.45\textwidth]{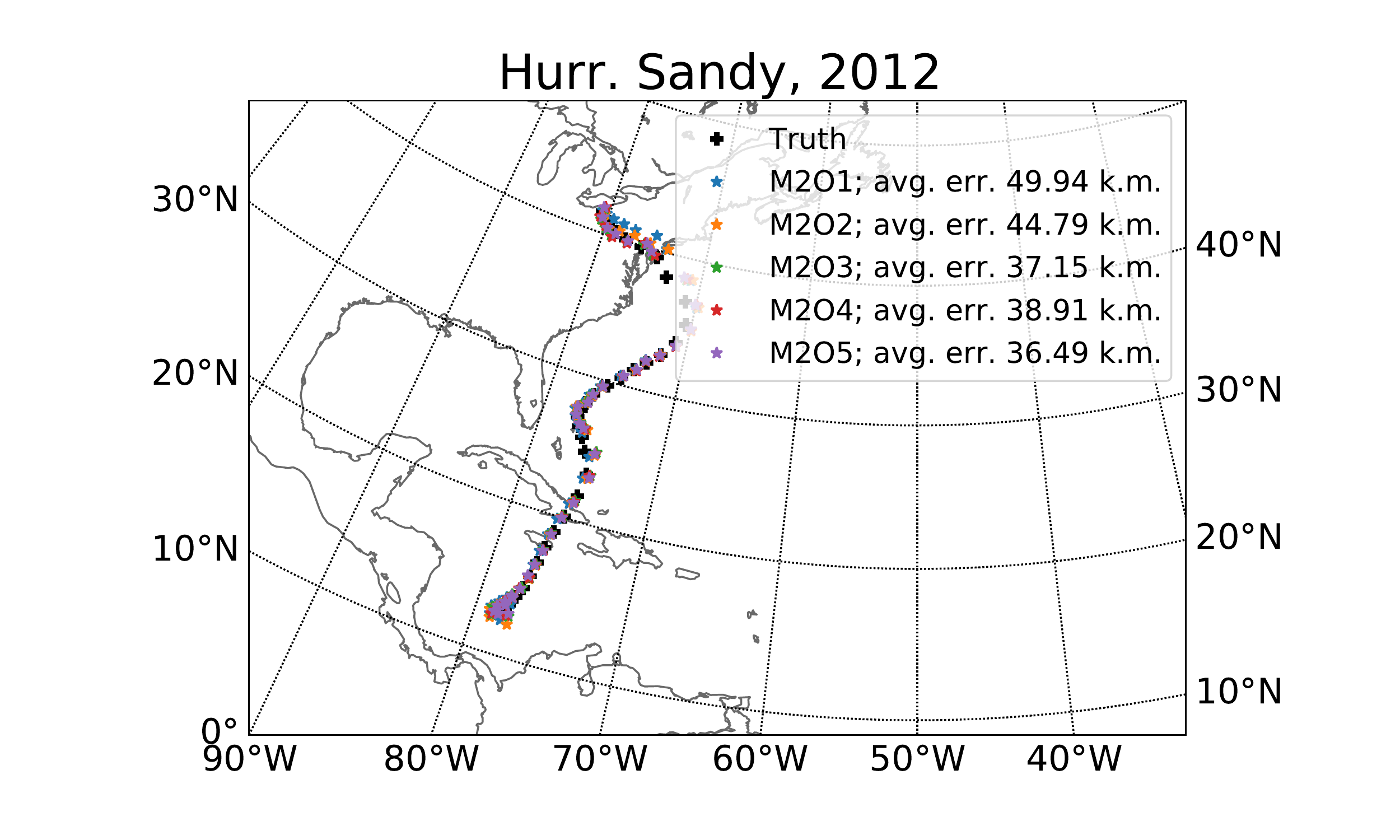}
        \includegraphics[trim=2cm 0cm 2cm 0cm,clip,width=0.45\textwidth]{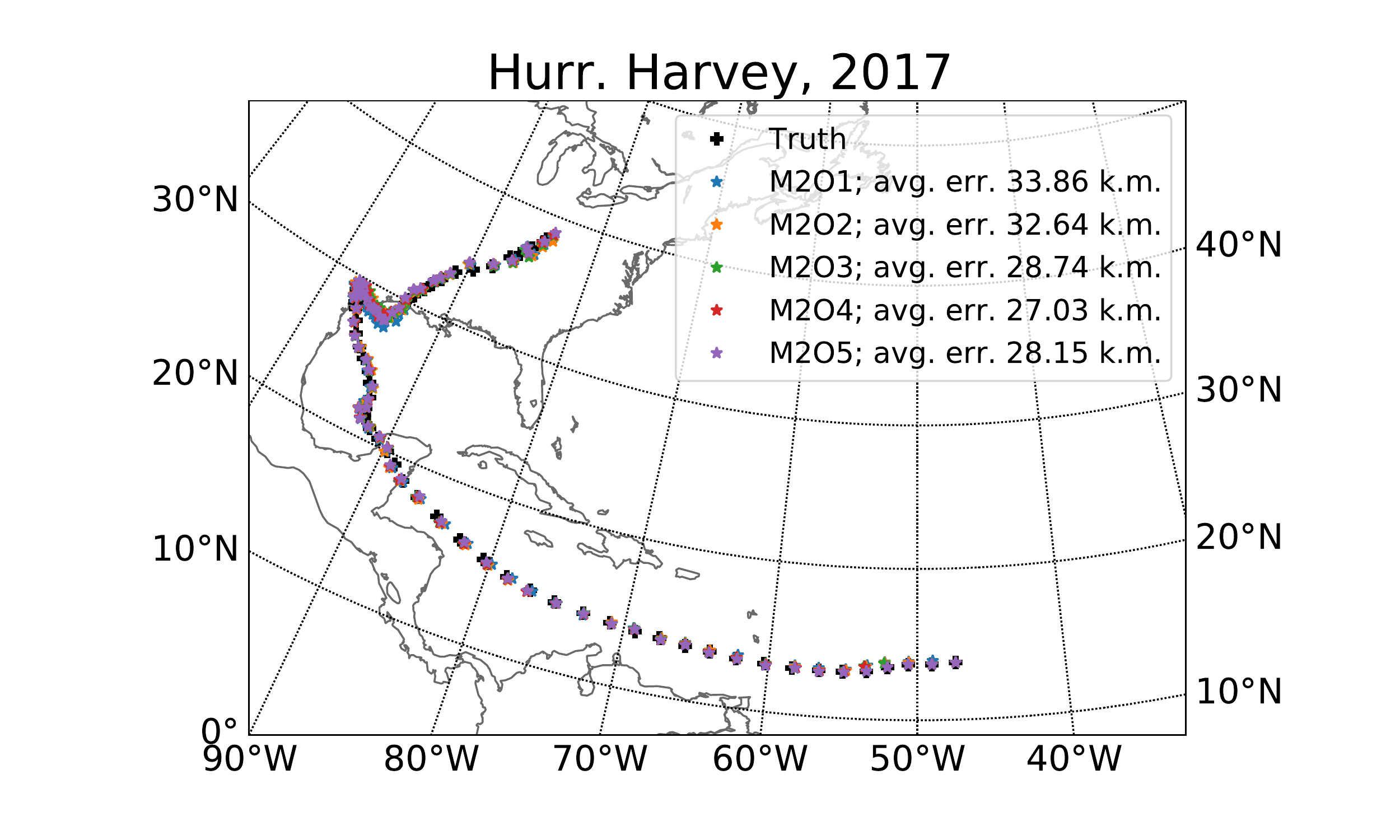}
        \caption{6--$hr$ forecast of select hurricane trajectories; Black `+' symbols represent the true locations; Colored `*' symbols represent the $M2On$ model forecasts. 
        Inputs to the model are the true feature data at each timestep. 
        Average error computed over the whole trajectory is also given. 
        }
        \label{fM2O-e}
    \end{center}
\end{figure*}

\begin{figure*}[!ht]
    \begin{center}
        \includegraphics[trim=2cm 0cm 2cm 0cm,clip,width=0.45\textwidth]{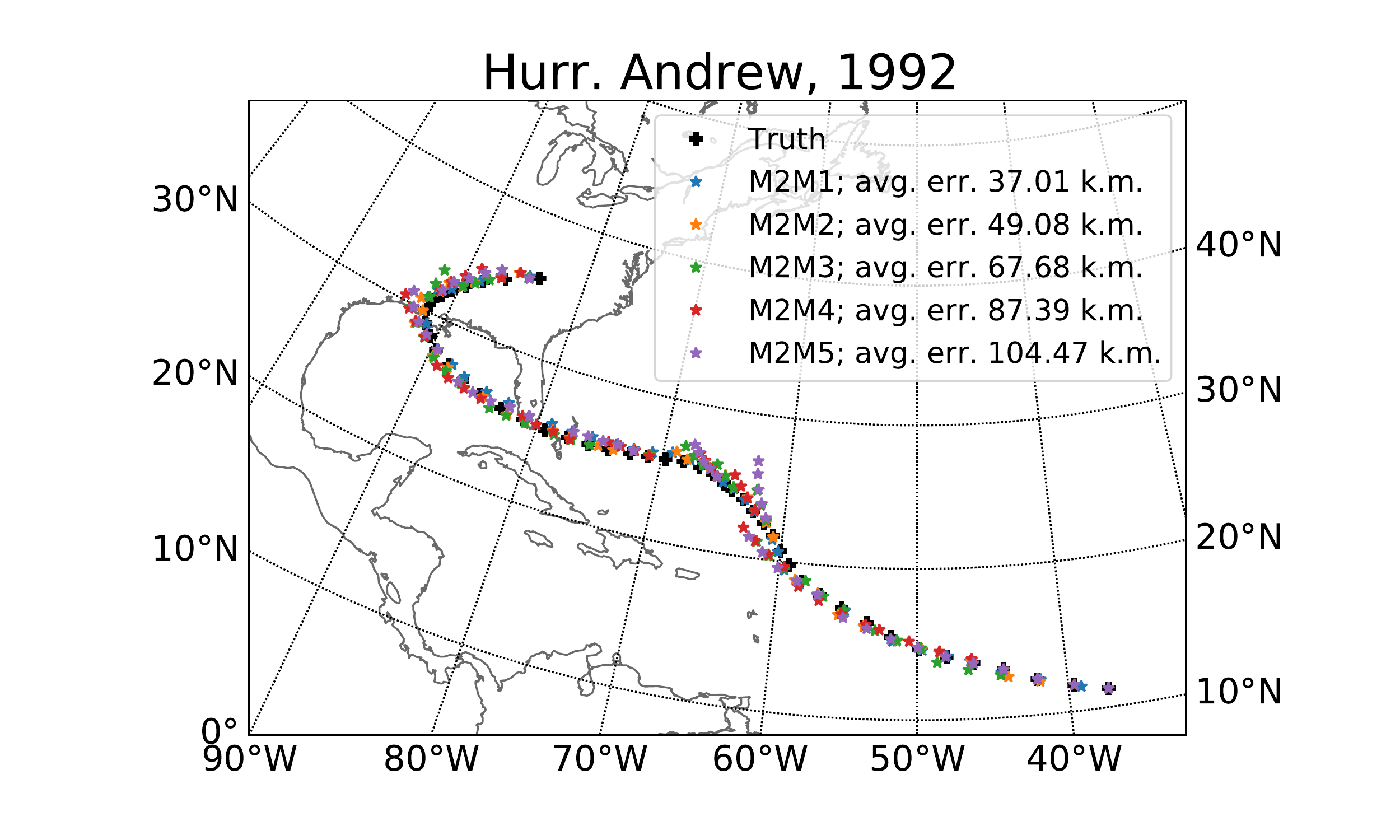}
        \includegraphics[trim=2cm 0cm 2cm 0cm,clip,width=0.45\textwidth]{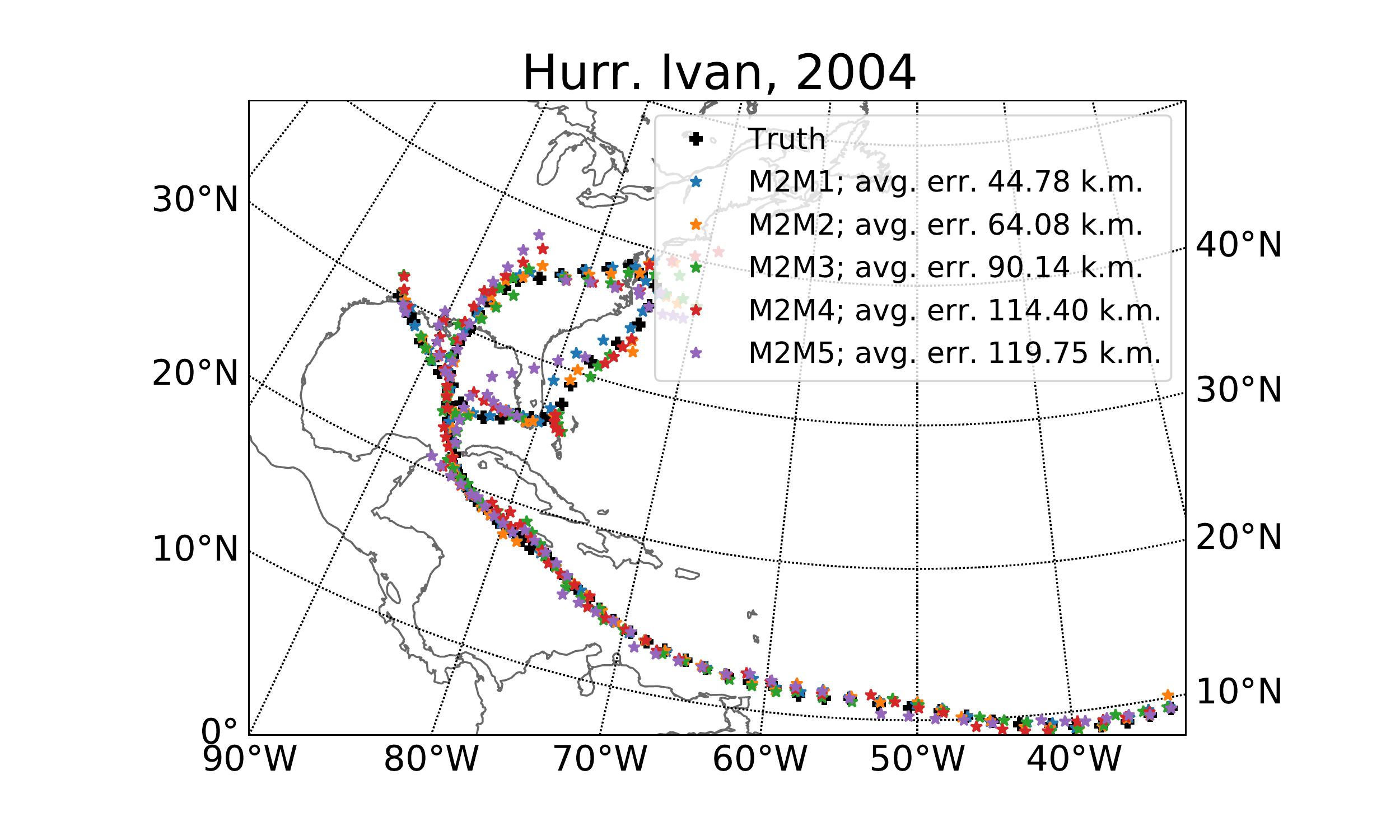}
        \includegraphics[trim=2cm 0cm 2cm 0cm,clip,width=0.45\textwidth]{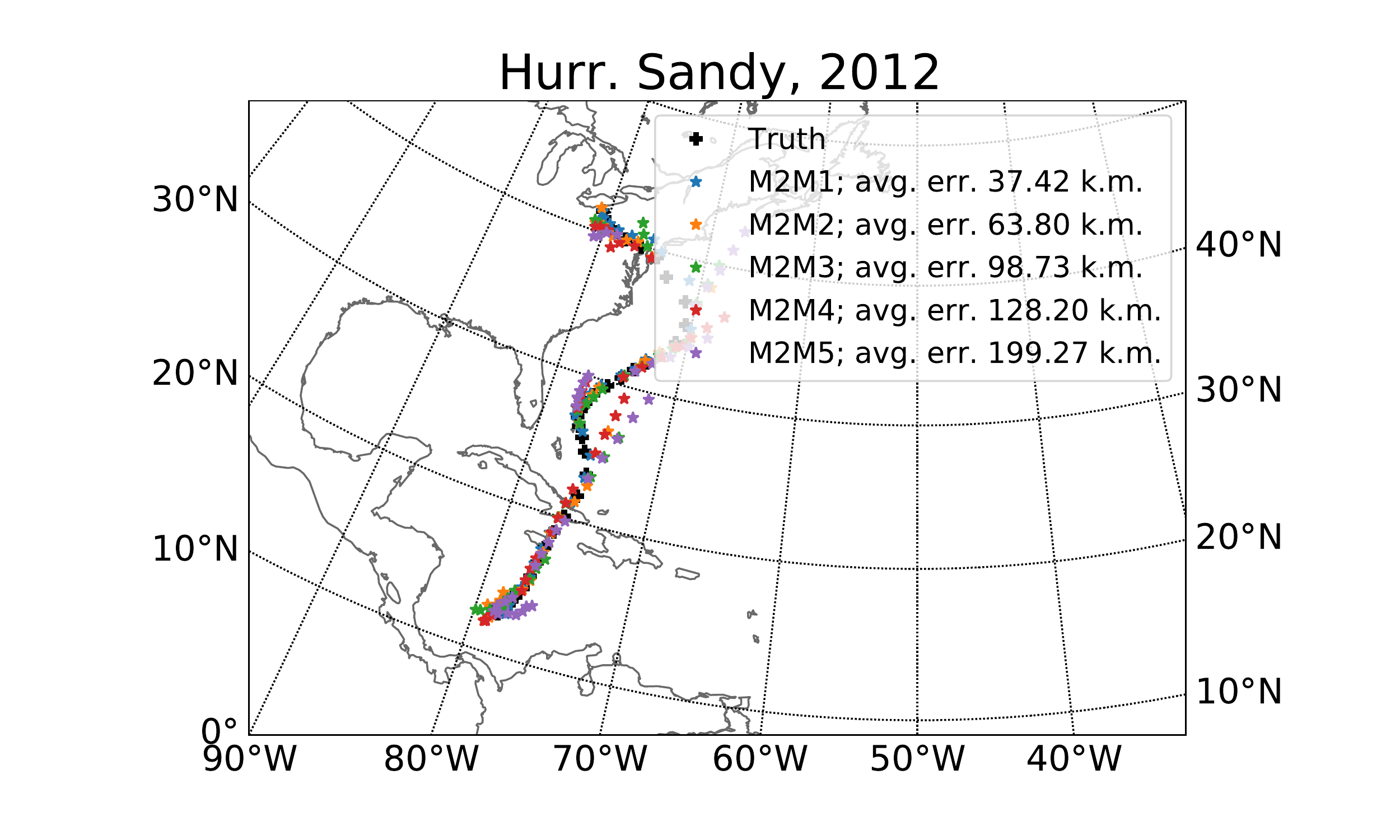}
        \includegraphics[trim=2cm 0cm 2cm 0cm,clip,width=0.45\textwidth]{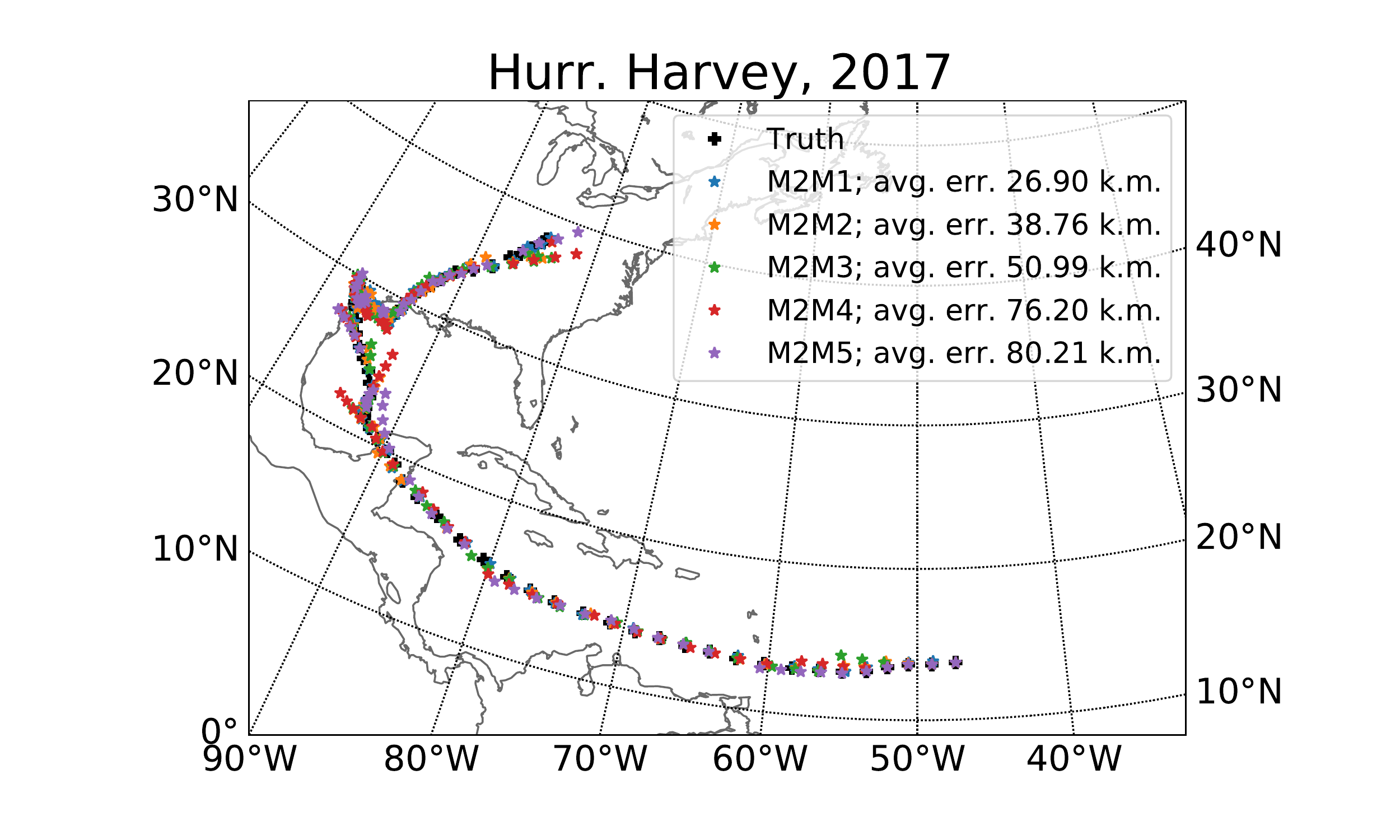}
        \caption{$6n-hr$ forecasts of select hurricane trajectories; Black `+' symbols represent the true locations; Colored `*' symbols represent the $M2Mn$ model forecasts. 
        Inputs to the model are the true feature data at each timestep. 
        Average error computed over the whole trajectory is also given.  
        }
        \label{fM2M-e}
    \end{center}
\end{figure*}

\subsection{Storm trajectory forecasting} 

Figures~\ref{f6hrErr} shows forecasting errors of the LSTM-RNN models computed over hundreds of validation and test storms. 
How do these models perform for individual test storms? 
This section reports trajectory predictions by these models for four extraordinarily powerful test storms: Andrew (1992), Ivan (2004), Sandy (2012) and Harvey (2017). 
Andrew, Ivan and Harvey were generated in the tropical Atlantic sub-basin; their trajectories exhibit the trend of trajectories of most storms generated in that sub-basin. 
Hurricane Ivan was chosen because it circled back across its original track near the end of its lifespan. 
The trajectory of hurricane Harvey has a sharp change toward south-east, opposite to the general trend. 
Hurricane Sandy was generated in the Caribbean sea. 
Its overall motion is consistent with the general trajectory trend of storms generated in the Caribbean. 
However, at the very beginning of its lifespan, it moves southward before turning sharply to travel northward. 
The aforementioned trajectory trends are complex compared to the general trends of storm motion. 
Therefore, these storms were chosen for demonstrating the performance of the models.

Figure~\ref{fM2O-e} shows the $6-hr$ forecasts by the $M2On$ models for the four hurricanes. 
Forecasting error was computed at all prediction instants and the mean forecasting error (per prediction step) was computed over the whole trajectories which are reported for all $M2O$ models. 
Exact features were used as inputs to the model. 
The $M2O1$ performs the worst model performs the worst with average errors of $\sim47$ $km$, $\sim52$ $km$, $\sim50$ $km$, and $\sim34$ $km$, for hurricanes Andrew, Ivan, Sandy, and Harvey, respectively. 
Model performance significantly improves for $n>2$. 
Among these three models, despite their complex trajectories, the maximum of the mean $6-hr$ forecasting errors computed for each of these storms were $\sim$ 28 $km$, 41 $km$, 39 $km$ and 29 $km$ for the hurricanes Andrew, Ivan, Sandy, and Harvey, respectively. 

Trajectories of the four aforementioned storms predicted by the $M2Mn$ models are shown in Figure \ref{fM2M-e}. 
To obtain predictions by the $M2Mn$ model for a given storm, exact input features at $n$ instants have been fed to the model to predict the storm locations at next $n$ time instants. 
So, the input time sequences are as $i = (0\to n-1), (n\to 2n-1), (2n\to 3n-1) ...$. 
Corresponding output sequences are $i = (n\to 2n-1), (2n\to 3n-1), (3n\to 4n-1) ...$. 
Therefore, each model predicts a target time record only once. 
Mean error incurred at each prediction step computed over the whole trajectory is reported for each model. 
For models with $n > 1$, the reported error for each model is the mean forecasting error per prediction, which is different from the mean $6n-hr$ forecast errors in Figure \ref{f6hrErr}($b$). 
Therefore, although the forecasting error for the $M2Mn$ model is highest at prediction step $n$, reported error in Figure \ref{f6hrErr}($b$) takes into account the model's more accurate predictions at the earlier prediction steps. 
The computed mean error per prediction step is smallest for Harvey and largest for Sandy. 
Mean errors computed over 12 and 24 hour forecasts for the $M2O2$ and $M2O4$ models for hurricane Harvey are 38.76 and 76.2 $km$. 
For hurricane Sandy these are $63.8$ and 128.2 $km$, respectively.

A hybrid approach is proposed for real-time storm evolution forecasting with the current models. 
When one record is available for a storm, $M2O1$/ $M2M1$ is used for prediction. 
When records of two time instants are available, a model trained on $n=2$ may be used. 
Similarly, models trained on increasing $n$ can be used, up to the point where 5 time records are available, beyond which the remaining span of the storm’s evolution is predicted by the model with $n=5$. 

\begin{table}
    \centering
    \resizebox{1.0\columnwidth}{!}{
        \begin{tabular}{ c c c c c } 
            \hline
            Time range & Forecast  & 2/3- probability circle radius & 2/3- probability circle radius \\
             & ($hrs$) & $M2M$ models ($nmiles$) & NHC ($nmiles$) \\
            \hline
            \multirow{2}{4em}{2005-2009} & 12 & 47 & 36 \\ 
                                         & 24 & 116 & 62 \\
            \hline
            \multirow{2}{4em}{2011-2015} & 12 & 42 & 8 \\ 
                                         & 24 & 112 & 23 \\                             
            \hline
            \multirow{2}{4em}{2016-2019} & 12 & 45 & 27 \\ 
                                         & 24 & 115 & 40 \\                                                          
            \hline
        \end{tabular}
        }
        \caption{Comparison of the $2/3$-probability circle radii corresponding to $12-$ and $24-hr$ forecast errors of the current $M2M$ models and the NHC in different time intervals.}
        \label{t1}
\end{table}

\subsection{Benchmarking}  

The literature on DL/ ML models applied to this problem does not use a fixed set of storms for comparison/ benchmarking. 
Rather, the NHC's forecasting error estimates given by the 2/3-probability circle radii for the periods 2005-2009, 2011-2015, and 2016-2019 are used. 
Models were trained keeping all storm events from these time frames in the test dataset as in Figure \ref{fDP}. 
The trained models were evaluated on test sequences generated from storms in those time periods only, and the respective 2/3-probability circle radii were computed (after sorting the errors in ascending order, errors for 2/3 of all sequences are less than the 2/3-probability circle radius). 
The radii are listed in table \ref{t1}. 
$M2M$ model forecasts are more consistent over different $5-yr$ periods than the NHC forecast errors. 
In addition, the 2/3-probability circle radii are competitive with the NHC's $12-hr$ forecast errors. 
In view of the small number of features it uses, the performance of the displacement probability approach is rather remarkable. 
Although the radius for the $24-hr$ forecasts is significantly larger for the $M2M$ models, interestingly, for the period 2016-2019, the mean and standard deviation of the $24-hr$ forecast errors for the $M2M5$ model are 198 $km$ 153 $km$, respectively. 
These are similar to the forecasting errors for the CLIPER statistical model used by the NHC (mean 201 $km$, std 149 $km$; see Table 4 of \citet{boussioux2020hurricane}). 

\subsubsection{Limitations}

\begin{figure}[!ht]
    \begin{center}
        \includegraphics[trim=1cm 0cm 0 0cm,clip,width=0.485\textwidth]{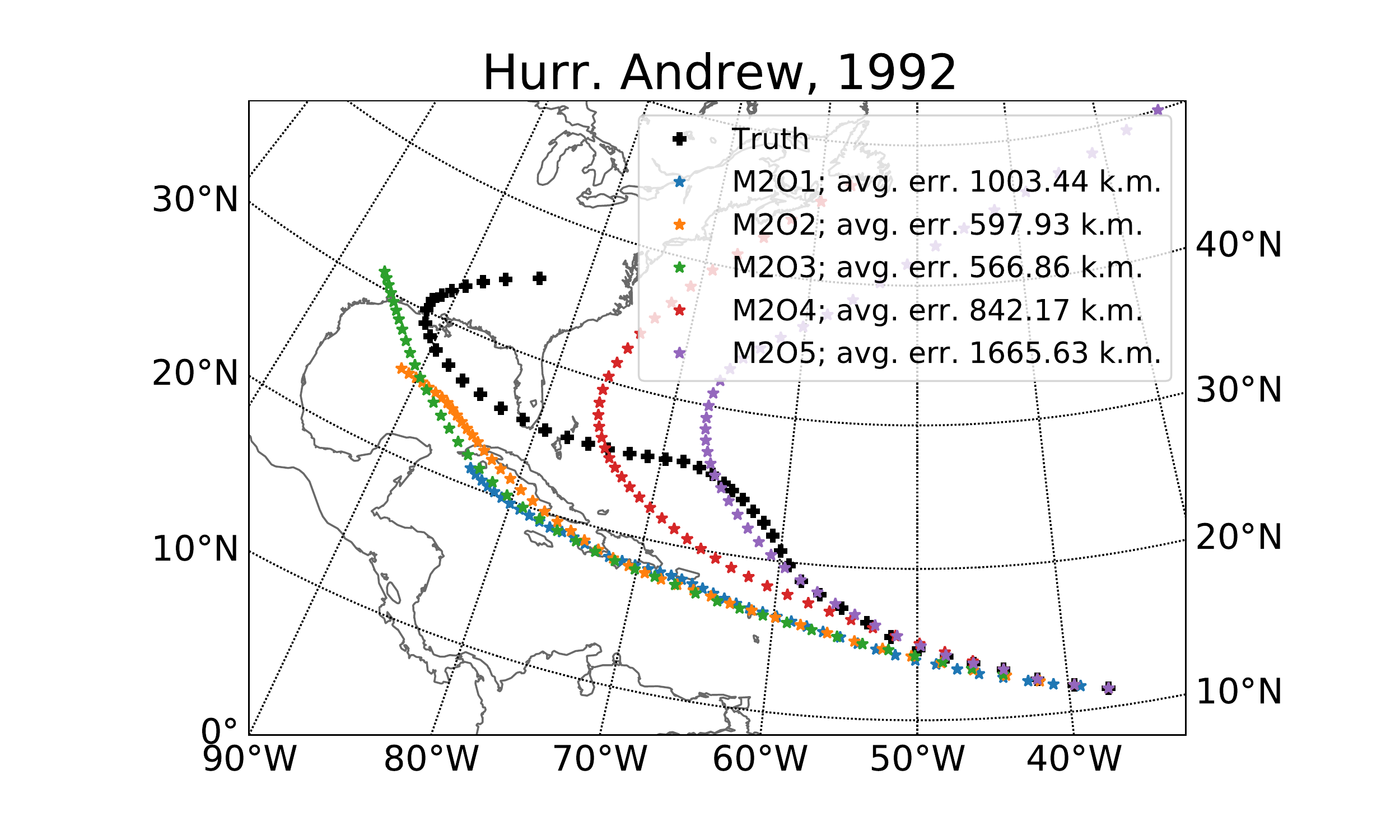}
        \includegraphics[trim=1cm 0cm 0 0cm,clip,width=0.485\textwidth]{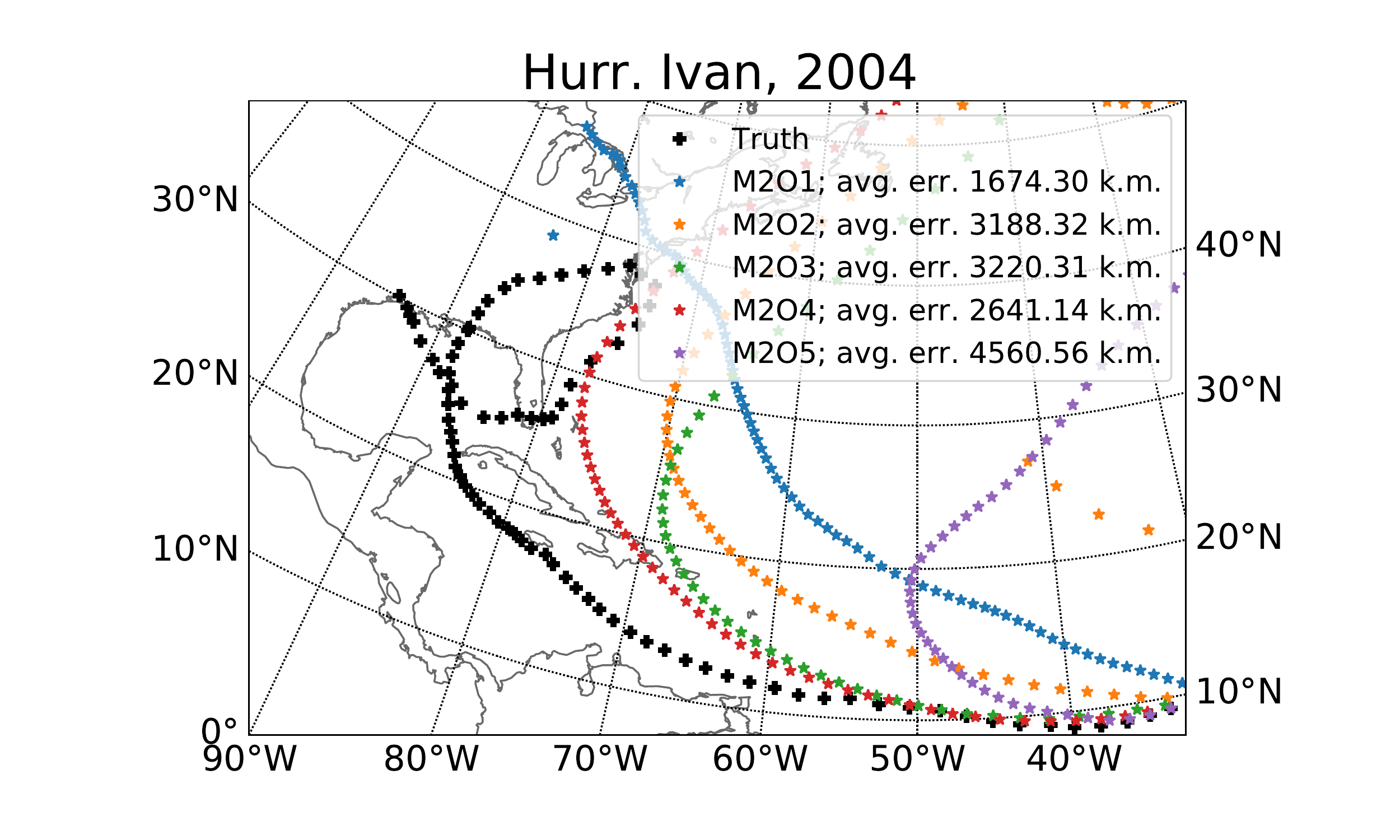}
        \includegraphics[trim=1cm 0cm 0 0cm,clip,width=0.485\textwidth]{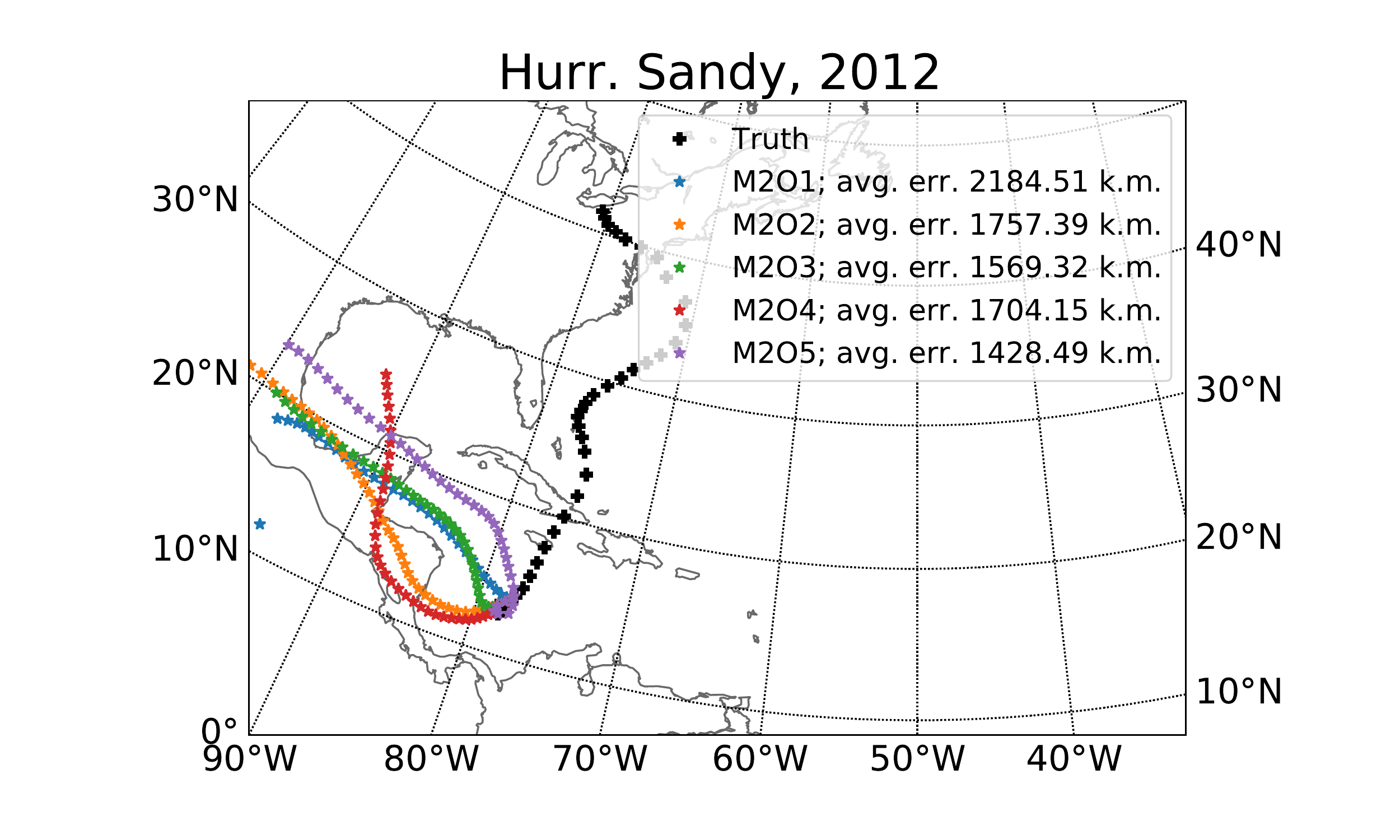}
        \includegraphics[trim=1cm 0cm 0 0cm,clip,width=0.485\textwidth]{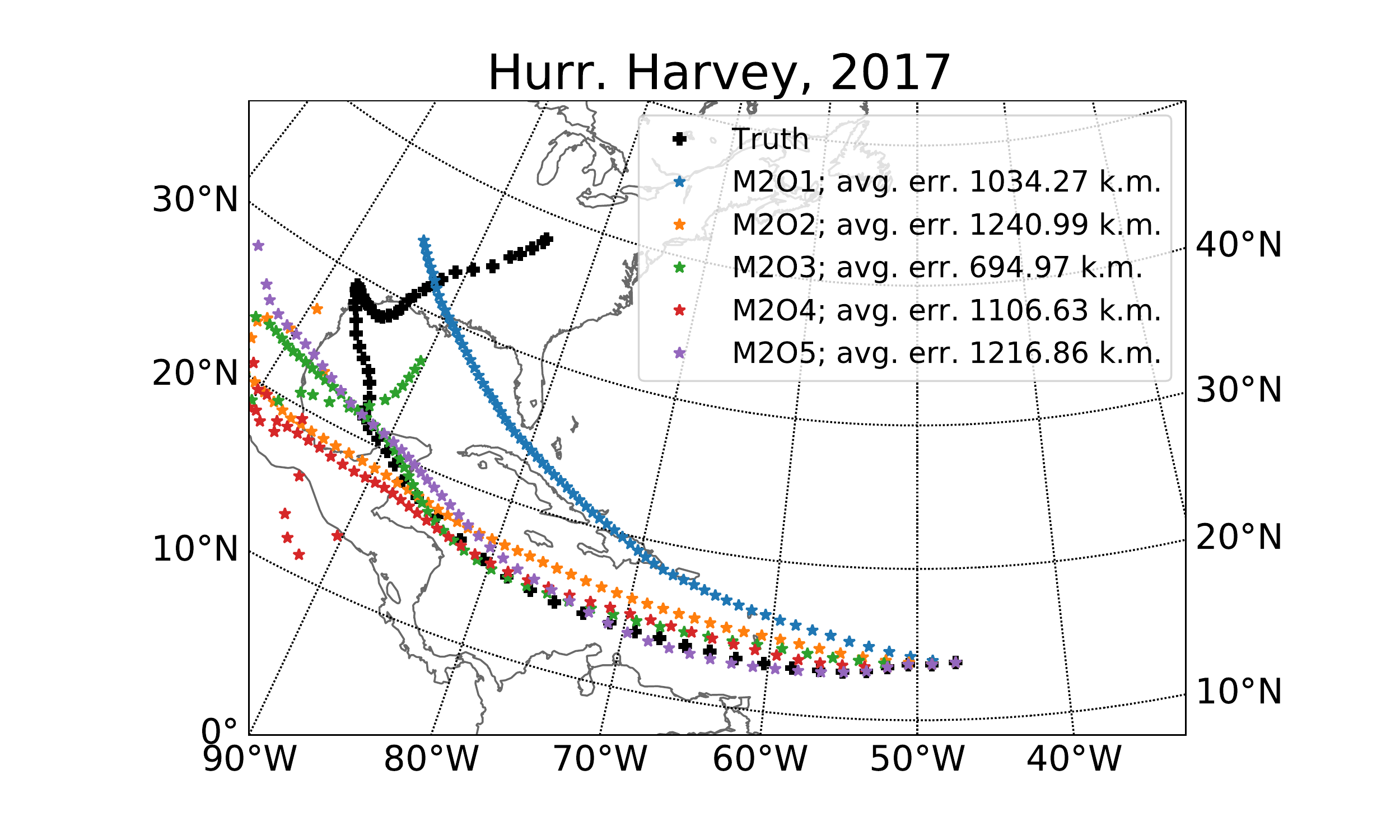}
        \caption{Forecast of whole trajectories of select hurricanes in advance given only the initial/ first sequence; actual trajectory is shown by the grey `+' symbols; the `*' symbols represent the predicted coordinates by the $M2On$ models. 
        Inputs to the models are updated based on predictions at previous prediction steps. 
        Average error in miles calculated over the whole trajectory is reported. 
        }
        \label{fM2O-p}
    \end{center}
\end{figure}

\begin{figure}[!ht]
    \begin{center}
        \includegraphics[trim=1cm 0cm 0 0cm,clip,width=0.485\textwidth]{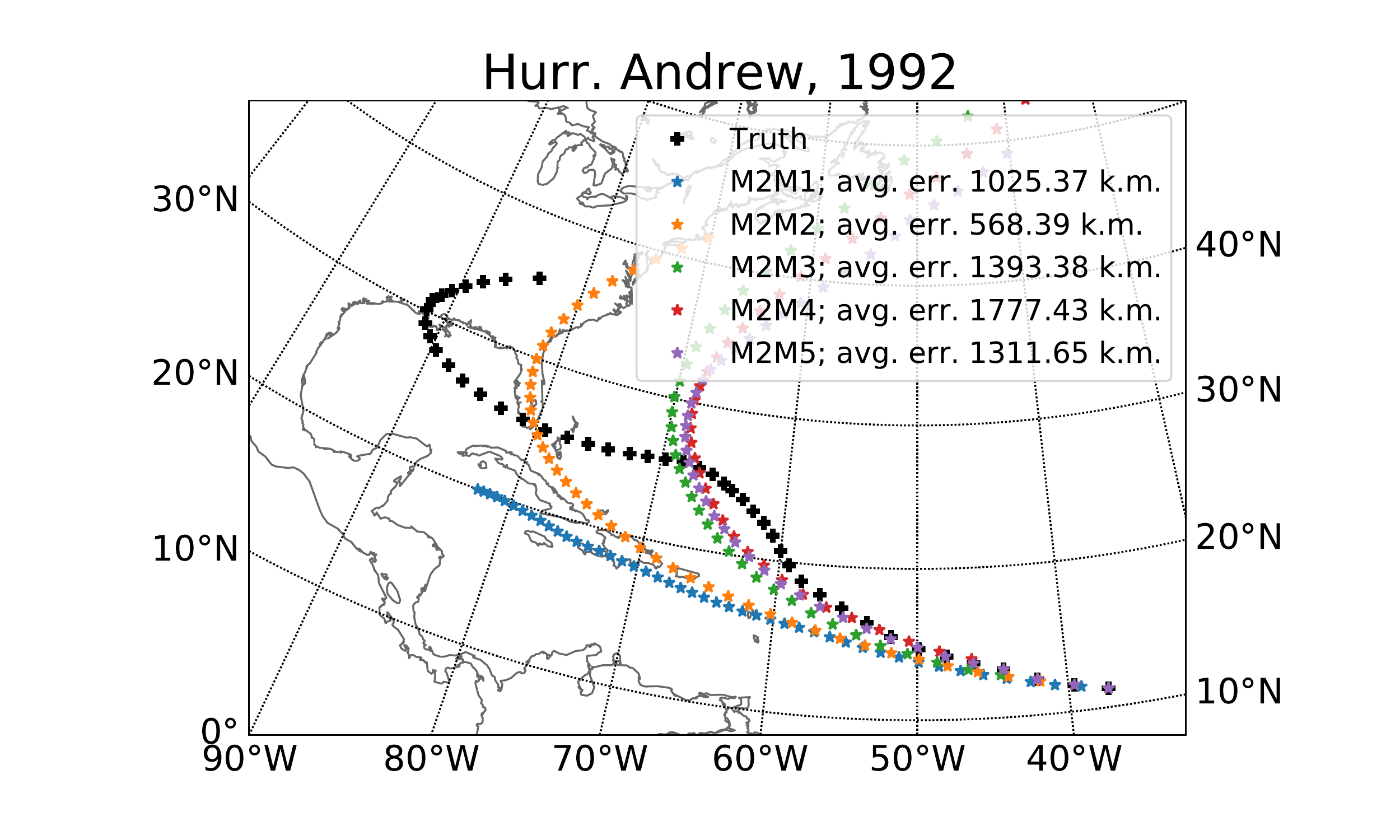}
        \includegraphics[trim=1cm 0cm 0 0cm,clip,width=0.485\textwidth]{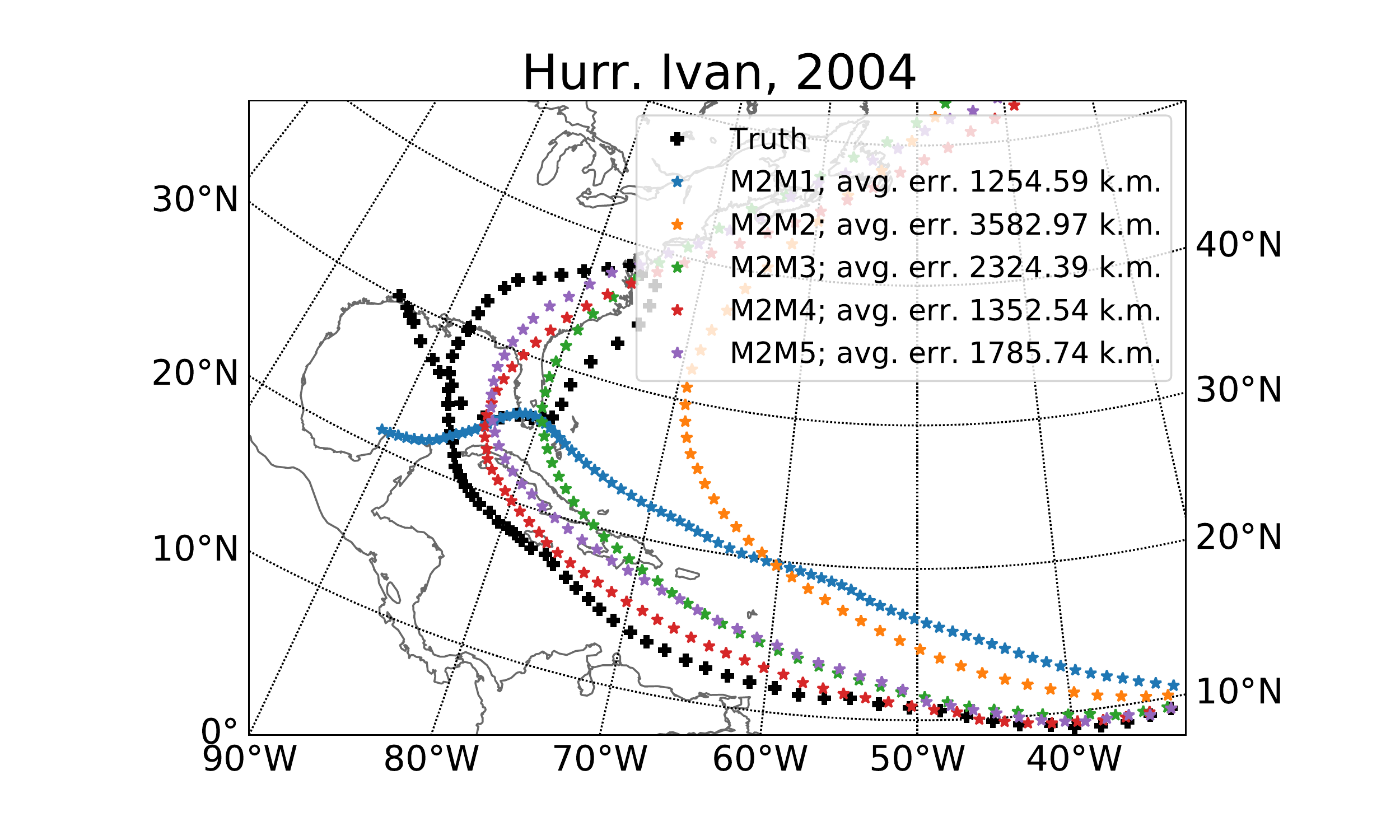}
        \includegraphics[trim=1cm 0cm 0 0cm,clip,width=0.485\textwidth]{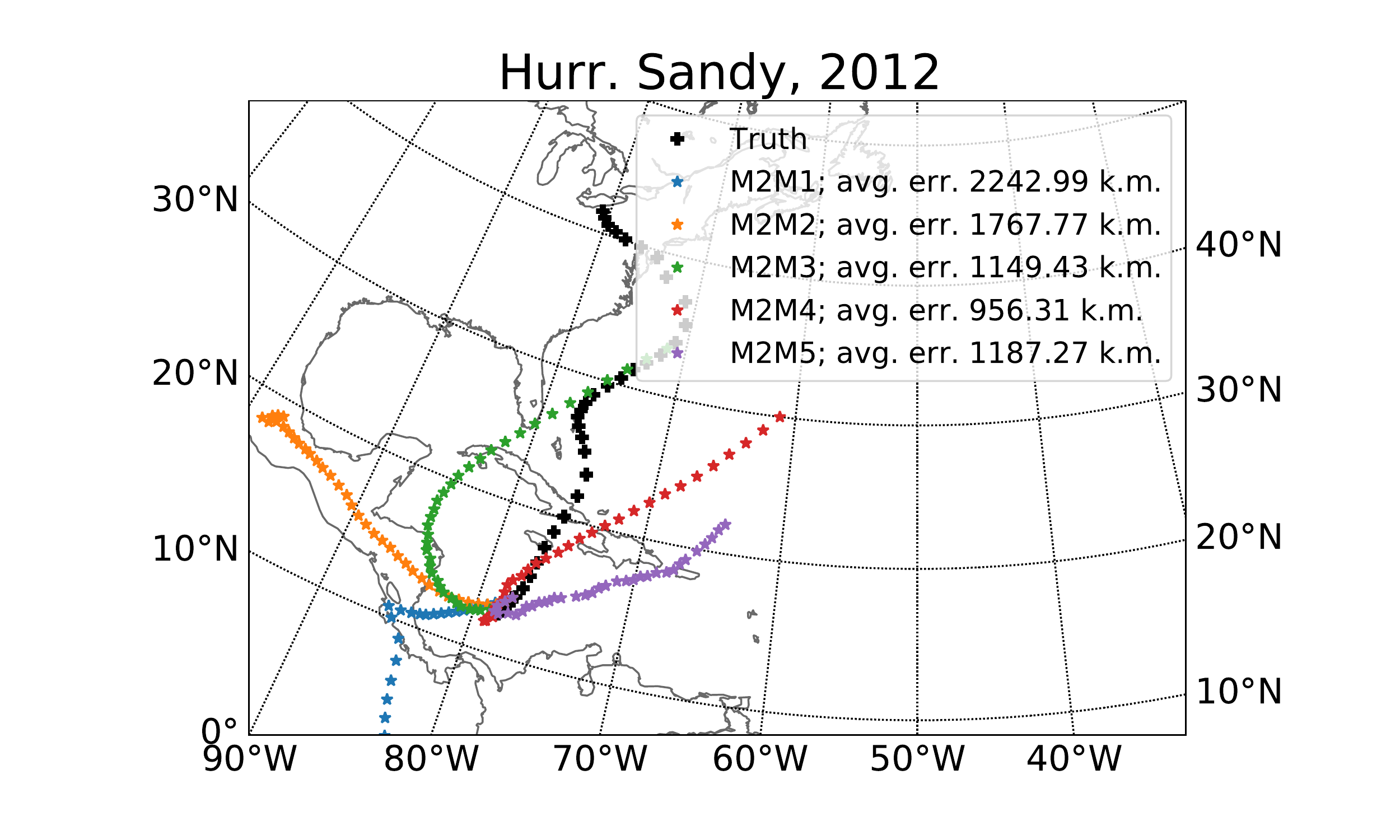}
        \includegraphics[trim=1cm 0cm 0 0cm,clip,width=0.485\textwidth]{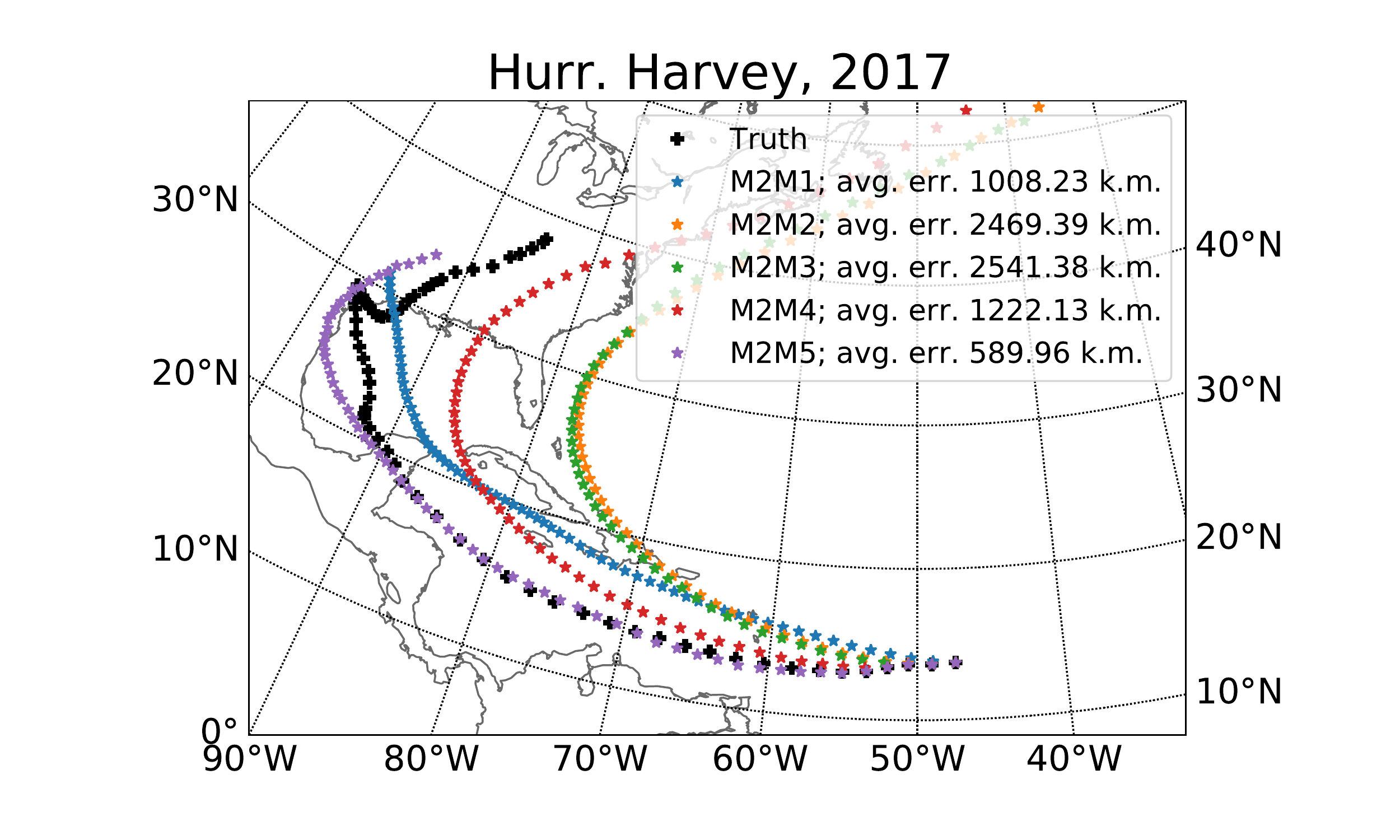}
        \caption{Forecast of whole trajectories of pre-selected hurricanes given only the initial/ first sequence; Black `+' symbols represent the true locations; Colored `*' symbols represent the $M2Mn$ model forecasts. 
        $2^{nd}$ prediction sequence onwards, inputs to the models are based on predictions at previous prediction steps. 
        Average error calculated over the whole trajectory is reported. 
        }
        \label{fM2M-p}
    \end{center}
\end{figure}

For real-time forecasting of storm trajectories, only the available time records from storm's current and past status may be used. 
Previous discussions on the performance of the LSTM-RNN models is based on a Machine/ Deep Learning perspective on the best/ ideal performance of the models for the problem being considered. 
The models were only fed the true features and predicted up to 30 hours in advance. 
However, longer forecasts of storm trajectories are typically necessary for disaster management purposes. 
To test these models' capabilities for this purpose, only the first input sequence of a storm's trajectory was fed to the model and the whole trajectory was predicted by the $M2Mn$ models. 
Inputs to the models from the second prediction sequence onwards include the predicted storm locations and corresponding derived storm speed, direction and displacement probabilities in previous time instants (only the true value for $w_m$ was used in all sequences). 
The trajectories for the four chosen test storms are shown in Figures~\ref{fM2O-p} and \ref{fM2M-p}. 
This is equivalent to a $228-hr$ forecast by the $M2M1$ model for hurricane Sandy, which had the shortest lifespan among the four test storms. 
Deviations of the predicted storm trajectories from the true trajectories in these Figures are basically illustrations of the compounded error growth previously discussed. 
Computed error in distance at each time step averaged over the whole trajectory is also reported for all the models.

The reported errors for the $M2Mn$ models are in most instances in thousands of kilometres for each predicted time step. 
The predicted trajectories in the initial stages of hurricane evolution are significantly closer to the true trajectories compared to later on in a storm's lifespan. 
This is to be expected, however, as the compounded errors incurred at each prediction step worsen the forecast at the next prediction step. 
Consequently, the predictions veer off from the true trajectories. 
None of the models is able to predict the complex loop in Ivan's trajectory. 
Similar deviations are obtained for the trajectory of hurricane Sandy. 
This storm initially travels southward and then turns about $180^\circ$ to travel northward. 
Although most models correctly predict the eventual northward motion of the storm at least qualitatively, the models are unable to predict the sudden change in direction of translation. 
The models perform better in predicting the trajectories of hurricanes Andrew and Harvey, which follow the general trajectory trends for the North Atlantic basin hurricanes. 
However, the forecast errors are still very large. 

The limitation of the LSTM-RNN models in predicting a whole storm trajectory from an initial condition is not surprising. 
Starting from its genesis, a storm's evolution is a complex nonlinear phenomenon dependent on several fluid and thermodynamic effects. 
This remains one of the most challenging problems in Geophysical Fluid Dynamics. 
For example, in \cite{leonardo2017verification}, testing several EPSs for forecasting North Atlantic hurricane trajectories, the minimum 24--$hr$ total track error (same as the error in distance used in the present work) increased by $\approx$ 75.64 -- 123.92 km for every $24-hr$ prediction interval for a $120-hr$ forecast for all storms in the period 2008 -- 2015. 
Mean $24-hr$ forecast errors for the test storms are $\approx 160$ $km$ for our $M2M4$ and $M2M5$ models (Figure~\ref{f6hrErr}). 
It should be noted that the EPSs are physics-based models that are guided by accurate fluid dynamic simulations, and therefore use a wealth of information unavailable to the current data-based models. 
Furthermore, following the current methodology, $M2Mn$ models with $n>5$ may be developed to further reduce long-term forecast errors due to compounded error accumulation associated with long-term trajectory forecasting. 

\section{Conclusions}
\label{sec5}

A family of LSTM-RNN models was developed to predict storm trajectories in the North Atlantic basin. 
Both Many-To-One and Many-To-Many type prediction architectures were used. 
The $6-hr$ storm d.p. feed the models with historical storm trajectory trends at a given location, which results in a significant gain in short-term forecasting accuracy (Figure \ref{fDP}). 
The proposed models are able to forecast up to 30 hours in advance. 
Mean forecasting errors computed over more than three hundred validation and test storms demonstrate the efficacy of these models. 
Overall, the $M2M$ models are more accurate. 
The minimum mean $6-hr$ forecasting error of all models was $\approx 30$ $km$. 
Considering that the storm-eye radii may extend up to 80 $km$ and storm radii may be as large as hundreds of kilometres \citep{willoughby2006parametric}, the model forecasts are reasonably accurate. 
$M2O$ models are more error-prone due to compounded error accumulation beyond $6-hr$ predictions; this limit for the $M2Mn$ model is $6n$ hours. 
Therefore, for long-term forecasting, the $M2M$ prediction models are more appropriate; these are more accurate at earlier prediction steps and the error increases linearly between prediction steps. 
The minimum and maximum mean $6-hr$ forecast errors for the $M2M$ models in predicting four chosen complex-trajectory test hurricanes were 27 and 37 $km$ (Figure \ref{fM2M-e}). 
The most significant advantage of the proposed models is their fast forecasting capability using modest computational resources. 
The model development methodology described herein uses readily available data and produces the requisite results in seconds. 
NHC uses an ensemble of models, including dynamical models that solve the fluid dynamics equations, that require hours of computation on supercomputers. 
A maximum of $30-hr$ storm evolution data is required. 
The models developed in this work are capable of providing predictions with as little as only one time record. 
The models presented herein are reliable and accurate, incurring errors comparable to those of state-of-the-art ensemble models used by the NHC for forecasting at least up to 12 hours (table \ref{t1}). 


  \bibliographystyle{elsarticle-num-names} 
  \bibliography{reference}





\end{document}